\documentclass[a4paper,11pt]{article}

\usepackage{jheppub}

\newcommand{\Pomeron}{I\!\!P}

\usepackage{epsfig}
\usepackage{graphicx}
\usepackage{amsmath}
\usepackage{hyperref}

\preprint{MS-TP-16-10}

\title{Diffractive dijet photoproduction in ultraperipheral collisions at the LHC in next-to-leading order QCD}

\author[a]{V. Guzey}
\author[b]{and M. Klasen}

\affiliation[a]{National Research Center ``Kurchatov Institute'',
Petersburg Nuclear Physics Institute (PNPI), Gatchina, 188300, Russia}

\affiliation[b]{Institut f\"ur Theoretische Physik, Westf\"alische Wilhelms-Universit\"at M\"unster,
Wilhelm-Klemm-Stra{\ss}e 9, D-48149 M\"unster, Germany}

\emailAdd{vguzey@pnpi.spb.ru}
\emailAdd{michael.klasen@uni-muenster.de}

\date{\today}

\abstract{We make predictions for the cross sections of diffractive dijet photoproduction
in $pp$, $pA$ and $AA$ ultraperipheral collisions (UPCs) at the LHC during Runs 1 and 2 using
next-to-leading perturbative QCD. We find that the resulting cross sections are sufficiently large and,
compared to lepton--proton scattering at HERA, have an enhanced sensitivity to small observed
momentum fractions in the diffractive exchange, commonly denoted $z_{\Pomeron}^{\rm jets}$,
and an unprecedented reach in the invariant mass of the photon--nucleon system $W$. We examine two competing schemes of diffractive QCD factorization breaking, 
which assume either a global suppression factor or a suppression for resolved photons only and
demonstrate that the two scenarios can be distinguished by the nuclear dependence of the
distributions in the observed parton momentum fraction in the photon $x_{\gamma}^{\rm jets}$.}

\begin{document} 
\maketitle
\flushbottom

\section{Introduction}
\label{sec:intro}

Ultraperipheral collisions (UPCs) of relativistic ions are characterized by large transverse distances (impact parameters) between the centers  of the colliding ions, exceeding the sum of their radii.
For such collisions, the strong interaction is suppressed and the ions interact electromagnetically 
through the emission of quasi-real photons~\cite{Fermi:1924tc,von Weizsacker:1934sx,Budnev:1974de}. 
The flux of these photons scales as 
$Z^2$, where $Z$ is the nuclear charge, 
and has a broad energy spectrum with the maximal photon energy in the laboratory frame scaling as $\gamma_L$,
where $\gamma_L$ is the nuclear Lorentz factor.
This allows one to study photon--photon and photon--nucleus scattering at unprecedentedly high energies~\cite{Baltz:2007kq}.

The UPC program at the LHC during Run 1 focused primarily on exclusive photoproduction of 
charmonia, in particular $J/\psi$ and $\psi(2S)$ mesons, which probes the gluon distribution of the target $g(x,\mu^2)$
at small values of the momentum fraction $x$ and a resolution scale $\mu^2={\cal O}(\rm few \ GeV^2)$~\cite{Ryskin:1992ui}.
This process was measured in  proton--proton ($pp$) collisions at 
$\sqrt{s_{NN}}=7$ TeV by the LHCb collaboration~\cite{Aaij:2013jxj,Aaij:2014iea},
 in proton--nucleus ($pA$) collisions at $\sqrt{s_{NN}}=5.02$ TeV by the ALICE collaboration~\cite{TheALICE:2014dwa}, and in
 Pb-Pb collisions at $\sqrt{s_{NN}}=2.76$ TeV by the ALICE collaboration~\cite{Abbas:2013oua,Abelev:2012ba,Adam:2015sia}
 ($ \sqrt{s_{NN}}$ is the invariant collision energy per nucleon).
 The analyses of these data at leading-order (LO) and next-to-leading order (NLO) QCD have provided new constraints on 
 the small-$x$ behavior of the gluon distribution in the proton 
$g_p(x,\mu^2)$ down to $x=6 \times 10^{-6}$~\cite{Jones:2013pga,Guzey:2013qza}
and of the gluon distribution in heavy nuclei $g_A(x,\mu^2)$ down to $x \approx 10^{-3}$~\cite{Adeluyi:2012ph,Guzey:2013xba}.
The data also 
restrict
the parameters and the strong interaction dynamics of the color 
dipole model approach~\cite{Lappi:2013am,Goncalves:2014wna} 
and the STARlight Monte Carlo generator~\cite{Klein:1999qj}.

The LHCb collaboration also measured exclusive photoproduction of $\Upsilon$ mesons in $pp$ UPCs at $\sqrt{s_{NN}}=7$ and 
8 TeV~\cite{Aaij:2015kea};
in perturbative QCD (pQCD), these data probe the proton gluon distribution at the resolution scale $\mu^2={\cal O}(\rm few \ tens \ GeV^2)$~\cite{Jones:2013pga}.
In addition, coherent photoproduction of $\rho$ mesons in nucleus--nucleus ($AA$) UPCs was measured by the STAR collaboration 
at RHIC at $\sqrt{s_{NN}}=64.4$, 130 and 200 GeV~\cite{Adler:2002sc,Abelev:2007nb,Agakishiev:2011me} 
and by the ALICE collaboration at $\sqrt{s_{NN}}=2.76$ TeV at the LHC~\cite{Adam:2015gsa}. 
These data probe the dynamics of soft high-energy $\gamma p$ and $\gamma A$ interactions, see e.g.\ Ref.\ \cite{Frankfurt:2015cwa}. 

Another potentially interesting process, which can be studied in $pp$, $pA$ and $AA$ UPCs at the LHC, 
is diffractive photoproduction of dijets, see Fig.~\ref{fig:photo_dijets}.
The measurement and the QCD analysis of this process in $pp$ and $pA$ UPCs will continue and extend the studies 
in lepton--proton scattering at HERA~\cite{Chekanov:2007rh,Aktas:2007hn,Aaron:2010su,Andreev:2015cwa,Klasen:2004qr,Klasen:2004ct,Klasen:2005dq,Klasen:2008ah,Klasen:2010vk},
giving a new handle on the key issue of factorization breaking and providing 
additional information on the proton diffractive PDFs. 
In $AA$ (and to some degree $pA$) UPCs at the LHC,
diffractive dijet photoproduction on nuclei presents an open field of research, 
which gives access to the novel unmeasured nuclear diffractive PDFs and the nuclear dependence
of factorization breaking.
Note that studies of diffractive dijet photoproduction in UPCs at the LHC are complimentary to those of diffractive dijet 
production in proton--antiproton scattering at the Tevatron~\cite{Affolder:2000vb,Klasen:2009bi} 
and in proton--proton scattering at the LHC~\cite{Aad:2015xis}.

The outline of this paper is as follows: Sections~\ref{sec:pp}, \ref{sec:pA} and \ref{sec:AA} contain our 
results for the $pp$, $pA$ and $AA$ cases, respectively, which we present in the same order.
First, we give the general expression for the cross section of diffractive dijet photoproduction in the considered UPC. 
Then we discuss its main ingredients, i.e.\ the photon flux, the rapidity gap survival probability and the diffractive parton distributions.
Third, we give and discuss our predictions for the cross sections of diffractive dijet photoproduction in UPCs in the LHC kinematics for Runs 1 and 2.
 In Sec.~\ref{sec:factor_br}, we discuss diffractive QCD factorization breaking and study its effect on our predictions.
Finally, we summarize our results in Sec.~\ref{sec:discussion}. For convenience, simple analytic fits to the suppression 
factors used in our calculations in the $pp$ and $pA$ cases are collected in the Appendix.

\section{Diffractive dijet photoproduction in proton--proton UPCs at the LHC}
\label{sec:pp}

\subsection{General expression for the cross section}
\label{subsec:pp_cs}

\begin{figure}[ht]
\begin{center}
\epsfig{file=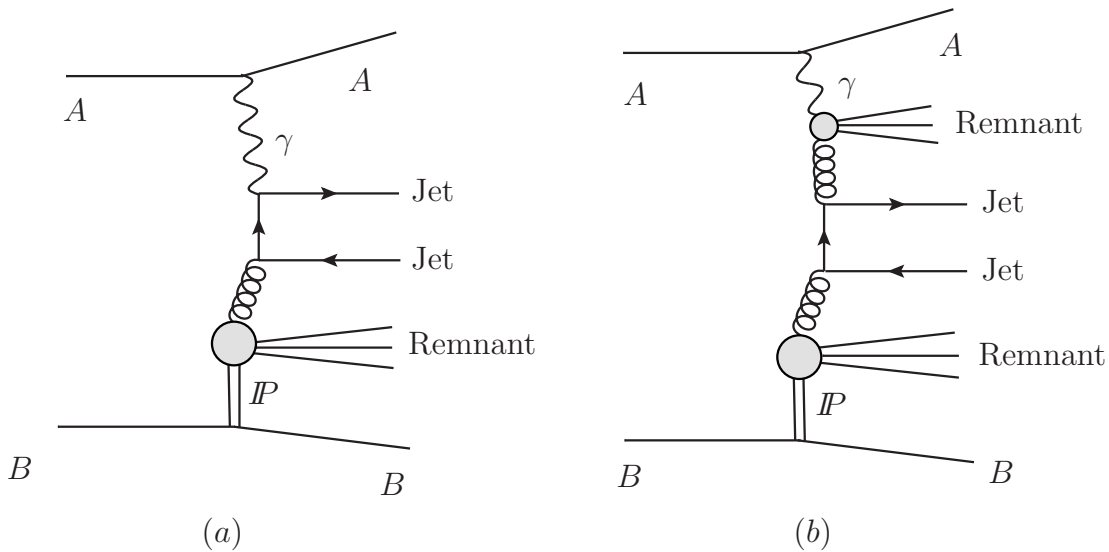,scale=0.85}
\caption{Typical leading-order Feynman graphs for diffractive dijet photoproduction in UPCs of 
hadrons $A$ and $B$.
Graphs $a$ and $b$ correspond to the direct and resolved photon contributions, respectively.}
\label{fig:photo_dijets}
\end{center}
\end{figure}

The mechanism of diffractive dijet photoproduction in ultraperipheral collisions of relativistic ions $A$ and $B$
is illustrated in Fig.~\ref{fig:photo_dijets}. The figure shows the dominant leading-order (LO)
Feynman graphs for the direct (graph $a$) and the resolved (graph $b$) photon contributions to the production of 
two quark jets.

Considering proton--proton UPCs, $A=B=p$, the cross section of diffractive dijet photoproduction can be written
as a sum of two terms:
\begin{eqnarray}
d \sigma(pp\to p+2{\rm jets}+X^{\prime}+Y) &=& d\sigma(pp\to p+2{\rm jets}+X^{\prime}+Y)^{(+)} \nonumber\\
&+& d \sigma(pp\to p+2{\rm jets}+X^{\prime}+Y)^{(-)} \,,
\label{eq:pp_inverse}
\end{eqnarray}
where $X^{\prime}$ stands for the produced diffractive final state $X$ after removing two jets and
$Y$ denotes the final state of the diffracting proton, which, besides the elastic state $Y=p$, may 
contain hadronic states with low invariant mass. Note that the possibility of the proton diffraction dissociation is
not explicitly shown in Fig.~\ref{fig:photo_dijets}.
The first and the second terms in Eq.~(\ref{eq:pp_inverse}) correspond to the diffracting proton moving along the positive and the negative $z$-axis, respectively. 
This reflects the ambiguity common for symmetric UPCs that either of the colliding ions can 
serve as a photon source and as a target~\cite{Baltz:2007kq}.
Since the jet pseudorapidities $\eta_1$ and $\eta_2$ are usually defined with respect to the
direction of the diffracting proton~\cite{Klasen:2002xb},
the two terms in Eq.~(\ref{eq:pp_inverse})
can be related to each other by inverting the sign of $\eta_1$ and $\eta_2$:
\begin{equation}
d \sigma(pp\to p+2{\rm jets}+ X^{\prime}+Y)^{(-)}=d \sigma(pp\to p +2{\rm jets}+ X^{\prime}+Y)_{|\eta_1 \to -\eta_1,\, \eta_2 \to -\eta_2}^{(+)} \,.
\label{eq:reverse_rule}
\end{equation}

The cross section $d \sigma(pp\to p+2{\rm jets}+ X^{\prime}+Y)^{(+)}$
can be readily written by analogy with the standard expression for the dijet diffractive photoproduction cross section
$d \sigma(ep\to e+2{\rm jets}+ X^{\prime}+Y)$ for lepton--proton scattering, see e.g.~\cite{Klasen:2002xb,Klasen:2004tza}:
\begin{eqnarray}
d \sigma(pp\to p &+&2{\rm jets}+X^{\prime}+Y)^{(+)}
=\sum_{a,b} \int_{t_{\rm cut}}^{t_{\rm min}} dt \int_{x_{\Pomeron}^{\rm min}}^{x_{\Pomeron}^{\rm max}} dx_{\Pomeron}
\int_0^1 dz_{\Pomeron} \int_{y_{\rm min}}^{y_{\rm max}} dy \int_0^1 dx_{\gamma}
\nonumber\\
&\times & S^2 (y) f_{\gamma/p}(y)  f_{a/\gamma}(x_{\gamma},\mu^2) f^{D(4)}_{b/p}(x_{\Pomeron},z_{\Pomeron},t,\mu^2)
d \hat{\sigma}_{ab \to {\rm jets}}^{(n)} \,,
\label{eq:pp_upc}
\end{eqnarray}
where $a$ and $b$ are parton flavors; $f_{\gamma/p}(y)$ is the flux of equivalent photons of the proton, 
which depends on the photon light-cone momentum fraction $y$; 
$f_{a/\gamma}(x_{\gamma},\mu^2)$ is the parton distribution function (PDF) of the photon, 
which depends on the parton light-cone momentum fraction $x_{\gamma}$ and 
the factorization scale $\mu$;  
$f^{D(4)}_{b/p}(x_{\Pomeron},z_{\Pomeron},t,\mu^2)$ is the diffractive PDF of the proton;
$d \hat{\sigma}_{ab \to {\rm jets}}^{(n)}$ is the elementary cross section for the production of an $n$-parton final state
in the interaction of partons $a$ and $b$; and the sum over $a$ involves quarks and gluons (the resolved
photon contribution) and the direct photon contribution with $a=\gamma$, which has support at LO only at $x_{\gamma}=1$.

In Eq.~(\ref{eq:pp_upc}), besides the standard expression for $ep$ scattering, 
we also explicitly introduced the rapidity gap survival factor of $S^2(y) \leq 1$, which takes into account 
the probability of soft inelastic interactions between the colliding protons, which populate, and thus suppress,
the final-state rapidity gaps. The factor of $S^2(y)$ depends on $y$ and the total invariant energy $\sqrt{s_{NN}}$;
in $pp$ UPCs, it can be viewed as a phenomenological factor modifying the photon flux 
$f_{\gamma/p}(y)$~\cite{Jones:2013pga,Schafer:2007mm}.

The QCD collinear factorization theorem for hard inclusive diffraction~\cite{Collins:1997sr} allows 
one to introduce universal diffractive PDFs  $f^{D(4)}_{b/p}(x_{\Pomeron},z_{\Pomeron},t,\mu^2)$ 
and to determine them by fitting to the measured diffractive
structure functions~\cite{Chekanov:2004hy,Aktas:2006hy,Aktas:2006hx}. 
The analysis also shows that for small values of $x_{\Pomeron}$,  $f^{D(4)}_{b/p}(x_{\Pomeron},z_{\Pomeron},t,\mu^2)$ 
can be written as the product of two factors~\cite{Ingelman:1984ns}:
\begin{equation}
f^{D(4)}_{b/p}(x_{\Pomeron},z_{\Pomeron},t,\mu^2)=f_{b/\Pomeron}(z_{\Pomeron},\mu^2) f_{\Pomeron/p}(x_{\Pomeron},t) \,,
\label{eq:Regge_fact}
\end{equation}
where $f_{b/\Pomeron}(z_{\Pomeron},\mu^2)$ is the PDF of the Pomeron (the lower blob in 
Fig.~\ref{fig:photo_dijets}) and $f_{\Pomeron/p}(x_{\Pomeron},t)$ is the Pomeron flux 
(the double line in Fig.~\ref{fig:photo_dijets}).
Note that the word ``Pomeron" here denotes the diffractive exchange.
Equation~(\ref{eq:Regge_fact}) helps to understand the meaning of the diffractive variables $z_{\Pomeron}$, $x_{\Pomeron}$ and 
$t$ entering Eq.~(\ref{eq:pp_upc}): $z_{\Pomeron}$ is the light-cone momentum fraction of a parton in the Pomeron;
$x_{\Pomeron}$ is the light-cone momentum fraction of the Pomeron in the proton; $t$ is the invariant momentum transfer squared.

In the measurements of diffractive dijet photoproduction in $ep$ scattering, the variables $y$, $x_{\Pomeron}$ and $t$ 
are directly reconstructed by detecting the scattered electron, the final proton and the diffractive final state, 
see e.g.~\cite{Aaron:2010su}:
\begin{eqnarray}
y & \equiv & \frac{q \cdot p}{k \cdot p}=1-\frac{E_e^{\prime}}{E_e} \,, \nonumber\\
x_{\Pomeron} & \equiv & \frac{ q \cdot (p-p_Y)}{q \cdot p}=\frac{E_X+P_{X,z}}{2 E_p}=\frac{M_X^2}{s y} \,, \nonumber\\
t & \equiv & (p-p_Y)^2 \,,
\label{eq:variables_diff}
\end{eqnarray}
where $p$, $p_Y$, $k$, and $q$ are the four-momenta of the initial proton, the final proton 
(with the possibility of diffraction dissociation into the state $Y$), the initial lepton and the photon, respectively;
$E_e$, $E_e^{\prime}$, and $E_X$ are the energies of the initial lepton, the final lepton, and the diffractive final state $X$,
respectively;
$P_{X,z}$ and $M_X$ are the $z$-component of the momentum and the invariant mass of the state $X$, respectively; and
$s=(k+p)^2$ is the square of the total center-of-mass energy of the collision.
In $ep$ scattering at HERA, 
the limits on $y$, $x_{\Pomeron}$ and $t$ 
are determined by the experimental conditions and cuts.
In contrast, in $pp$ UPCs the scattered protons travel along the beam pipe, and, hence, are undetected. 
As a result,  the limits on $y$, $t$ and $x_{\Pomeron}$ in Eq.~(\ref{eq:pp_upc}) are determined from 
general requirements to produce a diffractive final state.

The two remaining variables $z_{\Pomeron}$ and $x_{\gamma}$ in Eq.~(\ref{eq:pp_upc}) cannot be directly 
reconstructed by measuring the final state; their values can be compared to the hadron-level estimators
$z_{\Pomeron}^{\rm jets}$ and $x_{\gamma}^{\rm jets}$, respectively, which are reconstructed from the
measurement of the dijet and the diffractive final state~\cite{Aaron:2010su}:
\begin{eqnarray}
z_{\Pomeron}^{\rm jets} &=& \frac{\sum_{\rm jets} (E_i+P_{i,z})}{E_X+P_{X,z}} \,, \nonumber\\
x_{\gamma}^{\rm jets} &=& \frac{\sum_{\rm jets} (E_i-P_{i,z})}{E_X-P_{X,z}} \,,
\label{eq:variables_jets}
\end{eqnarray}
where the sum $\sum_{\rm jets}$ runs over the hadronic final states labeled ``$i$", which are included in the jets.

\subsection{Flux of equivalent photons in $pp$ UPCs}
\label{subsec:pp_flux}

The flux of quasi-real photons emitted by a relativistic proton (ion) can be found 
using the well-known Weizs{\"a}cker--Williams (WW) 
approximation~\cite{Fermi:1924tc,von Weizsacker:1934sx,Budnev:1974de,Bertulani:1987tz,Vidovic:1992ik}.
Since one also needs to take into account the charge and magnetization 
distribution in the proton,
in practical applications one often uses approximate expressions~\cite{Drees:1988pp,Kniehl:1990iv} 
reproducing the exact result with a few percent accuracy, see the discussion in Ref.\ \cite{Nystrand:2004vn}.

The photon flux produced by a relativistic charge $Z$ at the transverse distance $b$ from its center reads, see 
e.g.~\cite{Vidovic:1992ik}:
\begin{equation}
f_{\gamma/Z}(x,b)=\frac{\alpha_{\rm e.m.}Z^2}{\pi^2} \frac{1+(1-x)^2}{2x} \left|\int_0^{\infty}  \frac{dk_{\perp} k_{\perp}^2}{k_{\perp}^2 +
(x m_p)^2} F_{ch}(k_{\perp}^2+(x m_p)^2) J_1(b k_{\perp}) \right|^2 \,,
\label{eq:flux_vidovic}
\end{equation}
where $x$ is the fraction of the ion energy carried by the photon;
$\alpha_{\rm e.m.}$ is the fine-structure constant; $F_{ch}(k_{\perp}^2)$ is the charge form factor; 
$k_{\perp}$ is the photon  transverse momentum;
$J_1$ is the Bessel function of the first kind; and $m_p$ is the proton mass. 
Different expressions for the photon flux used in the literature correspond to various assumptions for $F_{ch}(k_{\perp}^2)$ 
and treatments of subleading terms in Eq.~(\ref{eq:flux_vidovic}).

The integration of $f_{\gamma/Z}(x,b)$ over all impact parameters gives the following general expression for the photon flux produced by a relativistic ion:
\begin{equation}
f_{\gamma/Z}(x)=\frac{\alpha_{\rm e.m.}Z^2}{\pi} \frac{1+(1-x)^2}{2x} \int_0^{\infty}  dk_{\perp}^2 k_{\perp}^2 \left(\frac{F_{ch}(k_{\perp}^2+(x m_p)^2)}{k_{\perp}^2 +(x m_p)^2}\right)^2   \,.
\label{eq:flux_vidovic_2}
\end{equation}
In our calculations of $pp$ UPCs we will use the result of Drees and Zeppenfeld (DZ)~\cite{Drees:1988pp}:
\begin{equation}
f_{\gamma/p}(x)=\frac{\alpha_{\rm e.m.}}{2 \pi} \frac{1+(1-x)^2}{x} \left[\ln A-\frac{11}{6}+\frac{3}{A}-\frac{3}{2 A^2}
+\frac{1}{3 A^3}\right] \,,
\label{eq:flux_p_DZ}
\end{equation}
where $A=1+(0.71 \ {\rm GeV}^2)/Q^2_{\rm min}$ and $Q^2_{\rm min}=(x m_p)^2/(1-x)$ is the minimal kinematically-allowed photon virtuality.
Alternatively, in the literature one also considers the photon flux 
produced by a relativistic point-like (PL) charge $Z$ passing 
a target at a minimum impact parameter $b_{\min}$:
\begin{equation}
f_{\gamma/Z}(x)=\frac{\alpha_{\rm e.m.} Z^2}{\pi} \frac{1}{x} 
\left[ 2 \zeta K_0(\zeta) K_1(\zeta)-\zeta^2 \left(K_1^2(\zeta)-K_0^2(\zeta)\right)\right] \,,
\label{eq:flux_proton_PL}
\end{equation}
where $\zeta=x m_p b_{\rm min}$ and $b_{\rm min}=0.7$ fm for the proton~\cite{Nystrand:2004vn}.

\begin{figure}[t]
\begin{center}
\epsfig{file=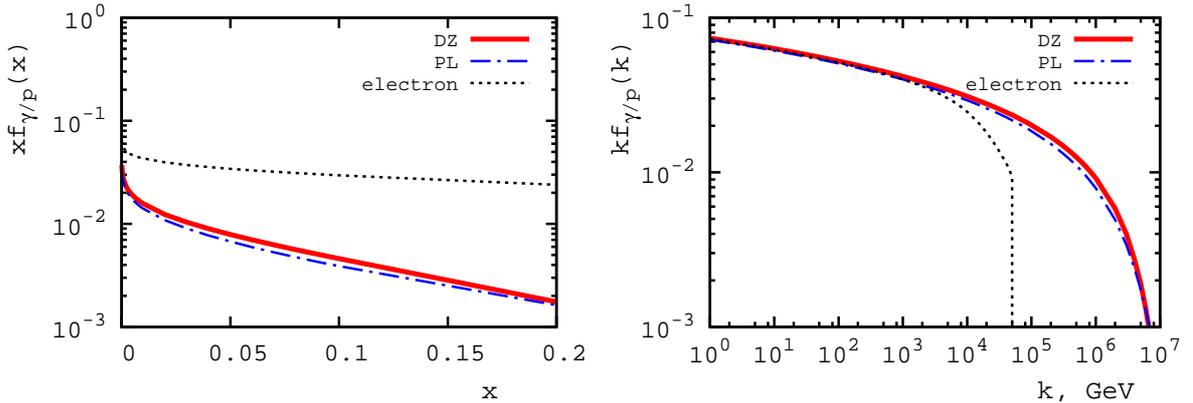,scale=1.25}
 \caption{Left: The proton and electron photon spectra $x f_{\gamma/p}(x)$ and $x f_{\gamma/e}(x)$, respectively, 
 as a function of the energy fraction $x$.
Right:  The photon spectra $k f_{\gamma/p}(k)$ and $k f_{\gamma/e}(k)$ as a function of the photon energy $k$ in 
the target rest frame.}
 \label{fig:photon_flux_Dec2015}
\end{center}
\end{figure}

It is illustrative to compare the photon flux of the proton with that of a relativistic electron~\cite{Budnev:1974de}:
\begin{equation}
f_{\gamma/e}(x)=\frac{\alpha_{\rm e.m.}}{2 \pi} \left[\frac{1+(1-x)^2}{x} \ln \frac{Q^2_{\rm max}(1-x)}{(m_e x)^2}+2 m_e^2 x
\left(\frac{1}{Q^2_{\rm max}}-\frac{1-x}{(m_e x)^2}\right)\right] \,,
\label{eq:flux_e}
\end{equation} 
where $m_e$ is the electron mass; $Q^2_{\rm max}$ is the maximal photon virtuality, which is usually 
determined by the experimental conditions.
Figure~\ref{fig:photon_flux_Dec2015} presents a comparison of the spectrum of equivalent photons of the proton
with that of the electron. In the left panel, we show $x f_{\gamma/p}(x)$ for the proton as a function of $x$ 
(the red solid and the blue dot-dashed curves corresponding to Eqs.~(\ref{eq:flux_p_DZ}) and (\ref{eq:flux_proton_PL}), respectively)
and $x f_{\gamma/e}(x)$ for the electron (the dotted black curve corresponding to Eq.~(\ref{eq:flux_e})).
 One can see from this panel that the energy spectrum for
the point-like electron is much flatter than that for the composite proton.
The photon energy $k$ scales as $\gamma_L m_p$ in the laboratory frame and as $2 \gamma_L^2 m_p$ in the target rest frame, 
where $\gamma_L$ is the Lorentz factor of the emitting ion. 
The right panel of Fig.~\ref{fig:photon_flux_Dec2015} shows the photon spectra $k f_{\gamma/p}(k)$ 
and $k f_{\gamma/e}(k)$
as a function of the photon energy $k$ in the proton target rest frame. 
For the proton beam, the curves correspond to proton--proton collisions at 
$\sqrt{s_{NN}}=7$ TeV. For the electron beam, the curve corresponds to the HERA kinematics with the 27.5 GeV electron beam and
the 920 GeV proton beam. One can see from the panel that $pp$ UPCs, in principle, allow one to obtain photon energies
exceeding those at HERA by two orders of magnitude.

We explained in Sec.~\ref{subsec:pp_cs} that the photon flux $f_{\gamma/p}(x)$ is somewhat 
reduced by the rapidity gap survival probability factor of $S^2(x)$. To estimate it, we use the method 
of Refs.~\cite{Jones:2013pga,Khoze:2000wk}, where $S^2(x)$ is calculated as a result of eikonalization of multiple Pomeron 
exchanges between the colliding protons:
\begin{equation}
S^2(x)=\frac{\int d^2 b \, |{\cal M}(x,b)|^2 P(s,b)}{\int d^2 b \, |{\cal M}(x,b)|^2} \,,
\label{eq:S2_pp}
\end{equation}
where $b$ is the impact parameter;
${\cal M}(x,b)$ is the diffractive amplitude of the process of interest in the impact parameter space;
$P(s,b)$ is the probability to not have the strong inelastic proton--proton interaction at the impact 
parameter $b$; and $s=s_{NN}$ for brevity.

The probability $P(s,b)$ in the two-channel eikonal model of Ref.~\cite{Khoze:2000wk} is 
\begin{eqnarray}
P(s,b) &=&\frac{1}{4 (1-\gamma^2)} \Big[(1+\gamma)^3 e^{-(1+\gamma)^2 \Omega(s,b)}+(1-\gamma)^3 e^{-(1-\gamma)^2 \Omega(s,b)}
\nonumber\\
&+& 2 (1-\gamma^2) e^{-(1-\gamma^2) \Omega(s,b)} \Big] \,,
\label{eq:P}
\end{eqnarray}
where $\gamma=0.4$ and $\Omega(s,b)$ is the proton optical density.
 Assuming that $\Omega(s,b)$  has the form of the effective Pomeron exchange trajectory, 
one obtains:
\begin{equation}
\Omega(s,b)=\alpha \frac{\sigma_{pp}^{\rm tot}(s)}{4 \pi B_P} e^{-b^2/(4 B_P)},
\label{eq:Omega}
\end{equation}
where $\sigma_{pp}^{\rm tot}(s)$ is the total proton--proton cross section; $B_P=B_0/2+\alpha^{\prime} \ln (s/s_0)$
is the slope of the $t$ dependence of the elastic $pp$ amplitude; and
the parameter $\alpha \geq 1$ results from eikonalized multiple Pomeron exchanges and is found 
from the requirement:
\begin{equation}
\sigma_{pp}^{\rm tot}(s)=2 \int d^2 b \left[1-\frac{1}{4} e^{-(1+\gamma)^2 \Omega(s,b)/2}-\frac{1}{4} e^{-(1-\gamma)^2 \Omega(s,b)/2}-\frac{1}{2} e^{-(1-\gamma^2) \Omega(s,b)/2}  \right] \,.
\label{eq:Omega_2}
\end{equation} 
 For $\sigma_{pp}^{\rm tot}(s)$, we use the fit
of the Review of Particle Physics~\cite{Agashe:2014kda}. The resulting values of $S^2(x)$ weakly depend
on the exact value of the slope $B_P$; in our calculation, we used $B_0=10$ GeV$^{-2}$ and $\alpha^{\prime}=0.25$ GeV$^{-2}$,
which correctly reproduce the slope of the elastic $pp$ cross section at small $|t|$~\cite{Khoze:2000wk}. 

For $|{\cal M}(x,b)|^2$ in Eq.~(\ref{eq:S2_pp}), we use the photon flux of the proton 
in the impact parameter space, Eq.~(\ref{eq:flux_vidovic}), with $F_{ch}(Q^2)=F_p(Q^2)=1/(1+Q^2/(0.71 \ {\rm GeV}^2))^2$.
 Since the contribution of small impact parameters $b < b_{\rm min} \approx 0.7$ fm to the photon flux, Eq.~(\ref{eq:flux_vidovic}), is small, 
 the integrand in Eq.~(\ref{eq:S2_pp}) receives the dominant
 contribution from $b > b_{\rm min}$, where $\Omega(s,b)$ is not large. As a result, one expects that the suppression
 due to $S^2(x)$ should not be large.

\begin{figure}[t]
\begin{center}
\epsfig{file=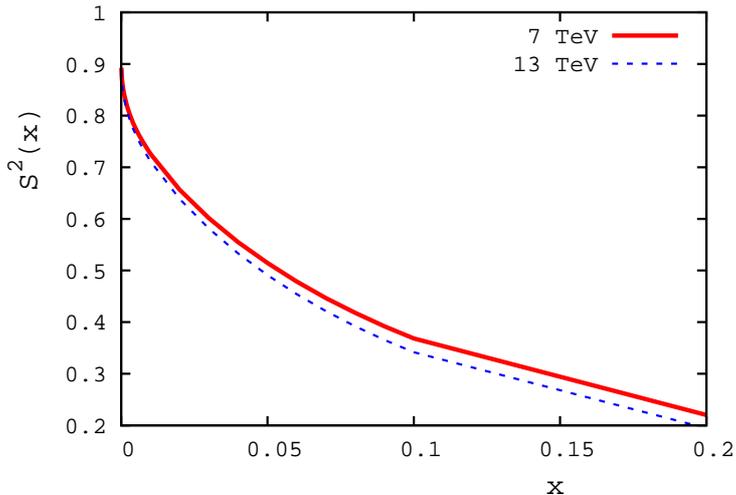,scale=1.1}
 \caption{The rapidity gap survival probability $S^2(x)$~(\ref{eq:S2_pp}) as a function of the photon momentum fraction $x$
 for $\sqrt{s_{NN}}=7$ TeV (red solid curve) and $\sqrt{s_{NN}}=13$ TeV (blue dashed curve).}
 \label{fig:S2_pp}
\end{center}
\end{figure}

The resulting values of $S^2(x)$ as a function of the photon light-cone momentum fraction $x$ 
are shown in Fig.~\ref{fig:S2_pp}. The red solid curve corresponds to $\sqrt{s_{NN}}=7$ TeV, and the blue dashed curve is for
 $\sqrt{s_{NN}}=13$ TeV. Since larger values of $x$ correspond effectively to smaller values of the impact parameter $b$, where 
 the suppression due to the strong interaction is stronger, $S^2(x)$ decreases with an increase of $x$. 
 Also, the suppression somewhat increases with an increase of the invariant collision energy $\sqrt{s_{NN}}$ as a result of the 
 increase of $\sigma_{pp}^{\rm tot}(s)$ and the optical density $\Omega(s,b)$.
 For convenience, a simple fit to the curves in Fig.~\ref{fig:S2_pp} is given in the Appendix.
Note that our values of $S^2(x)$ are consistent with the results for $S^2(x)$ for exclusive photoproduction of $J/\psi$ and 
$\Upsilon$ mesons in $pp$ UPCs~\cite{Jones:2013pga} 
and also with the results of the calculation using a simpler model for the rapidity gap survival of 
Ref.~\cite{Frankfurt:2006jp}.

\subsection{Results}
\label{subsec:pp_results}

We performed next-to-leading order (NLO) pQCD calculations implementing the inclusive $k_{T}$-cluster algorithm~\cite{Klasen:2010vk} of the cross section of diffractive photoproduction of dijets in $pp$ UPCs, Eq.~(\ref{eq:pp_inverse}), using the following cuts (compare to the cuts used, e.g., in Ref.~\cite{Aaron:2010su}):
\begin{eqnarray}
&&  0 < y < 1 \,, \nonumber\\
&& E_{T}^{\rm jet 1} > 20 \ {\rm GeV} \,, \quad \quad E_{T}^{\rm jet 2} > 18 \ {\rm GeV} \,, \nonumber\\
&& -6 < \eta^{\rm jet 1,2} < -6\,, \nonumber\\
&& x_{\Pomeron} \leq 0.03 \,, \quad z_{\Pomeron}^{\rm jets} \leq 1 \,, \nonumber\\
&& M_Y \leq 1.6 \ {\rm GeV} \,, \quad |t| < 1 \ {\rm GeV}^2 \,,
\label{eq:UPC_lj}
\end{eqnarray}
where $E_{T}^{\rm jet 1,2}$ and $\eta^{\rm jet 1,2}$ are the transverse energies and the pseudorapidities of 
the two jets, respectively.
For input, we used the DZ photon flux of the proton, Eq.~(\ref{eq:flux_p_DZ}), modified by the rapidity gap survival probability 
$S^2(y)$, Eq.~(\ref{eq:S2_pp}), the GRV-HO photon PDFs~\cite{Gluck:1991jc}, and the 2006 H1 proton diffractive PDFs 
(fit B)~\cite{Aktas:2006hy}. 

\begin{figure}[t]
\begin{center}
\epsfig{file=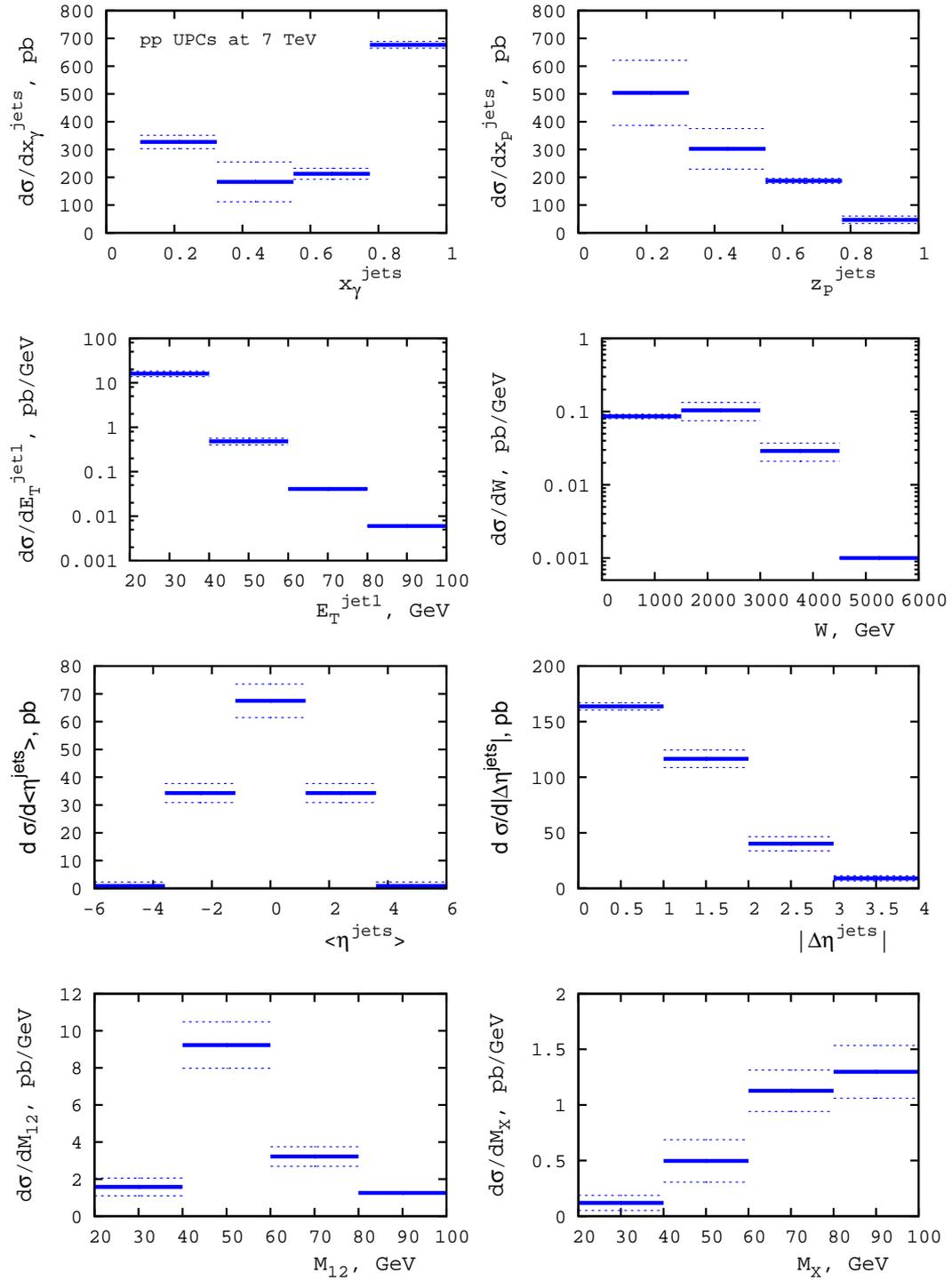,scale=1.1}
 \caption{The differential cross section of diffractive photoproduction of dijets $d \sigma(pp\to p+2{\rm jets}+ X^{\prime}+Y)$
 as a function of various variables in $pp$ UPCs at $\sqrt{s_{NN}}=7$ TeV.}
 \label{fig:upc_pp_scale_S2}
\end{center}
\end{figure}

\begin{figure}[t]
\begin{center}
\epsfig{file=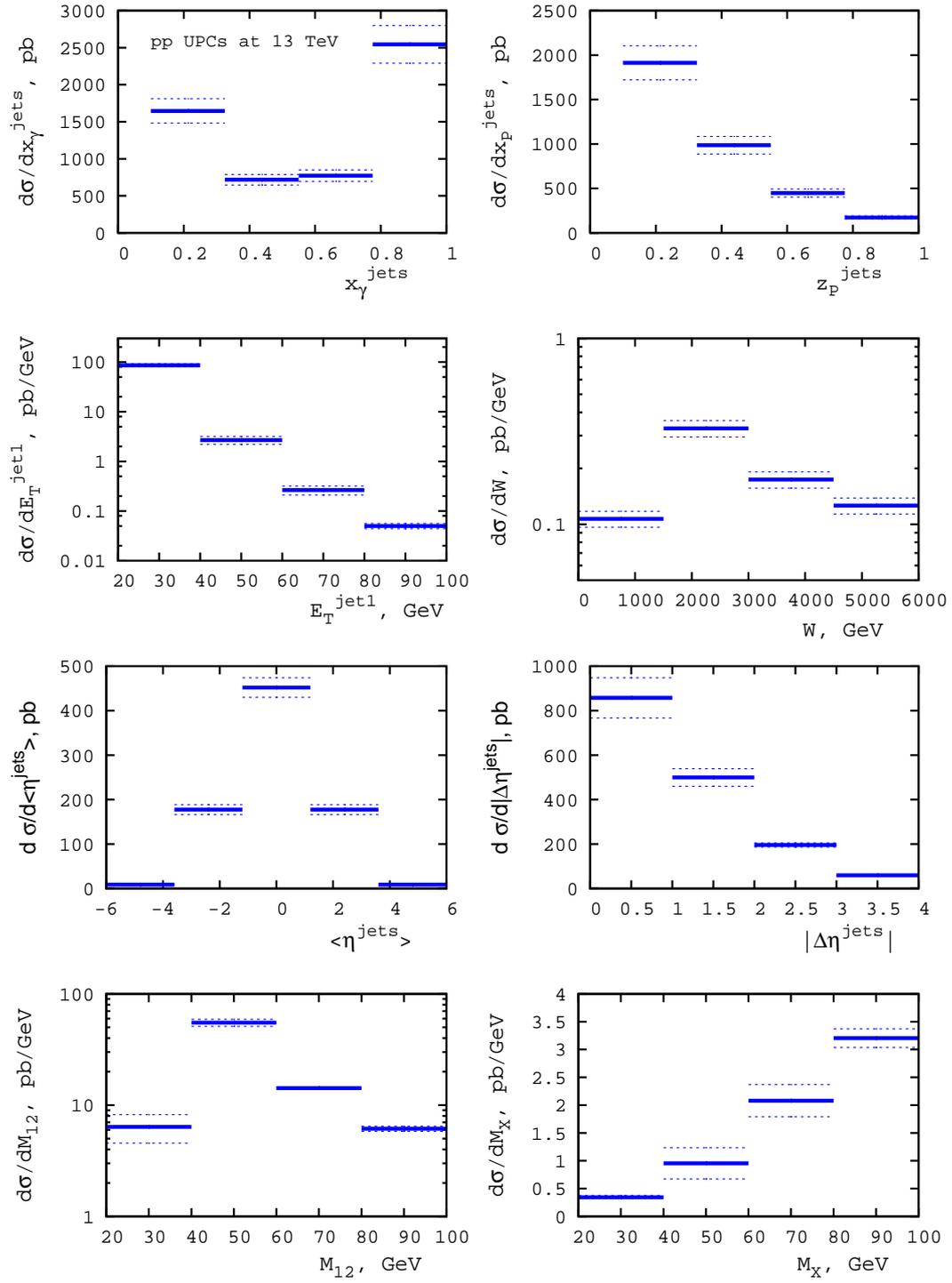,scale=1.1}
 \caption{The same as in Fig.~\ref{fig:upc_pp_scale_S2}, but at $\sqrt{s_{NN}}=13$ TeV.}
 \label{fig:upc_pp_scale_13_S2}
\end{center}
\end{figure}

Figures~\ref{fig:upc_pp_scale_S2} and \ref{fig:upc_pp_scale_13_S2} show
the resulting diffractive dijet photoproduction cross sections for $pp$ UPCs at $\sqrt{s_{NN}}=7$ TeV and $\sqrt{s_{NN}}=13$ TeV, respectively. Different panels present the cross section as a function of 
$x_{\gamma}^{\rm jets}$, $z_{\Pomeron}^{\rm jets}$, 
$E_T^{\rm jet 1}$, the invariant mass of the photon--proton system $W$, $\langle \eta^{\rm jets} \rangle=(\eta_1+\eta_2)/2$, $|\Delta \eta^{\rm jets}|=|\eta_1-\eta_2|$, the invariant mass of the dijet system 
$M_{12}$, and $M_X$, see~\cite{Aaron:2010su}.
The central thick solid lines show the result of the calculation, when the renormalization and factorization scale $\mu$ is identified 
with the transverse energy of jet 1, $\mu=E_{T}^{\rm jet 1}$; the dotted lines correspond to the calculation with 
$\mu=2 E_{T}^{\rm jet 1}$ and $\mu=(1/2) E_{T}^{\rm jet 1}$. Thus, the spread between the dashed lines quantifies the 
theoretical uncertainty of our NLO calculations associated with the choice of the factorization and renormalization scales.
One can see from the figures that for our choice of $E_T^{\rm jet 1}> 20$ GeV, this uncertainty is rather insignificant.

Several features of the presented results merit discussion and comparison to
diffractive dijet photoproduction in $ep$ scattering at HERA~\cite{Chekanov:2007rh,Aktas:2007hn,Aaron:2010su}.
First, the predicted yields are comparable to those observed in the $ep$ case; the integrated cross section is ${\cal O}(\rm hundreds \ pb)$ at $\sqrt{s_{NN}}=7$ TeV and
 ${\cal O}(\rm few \ nb)$ at $\sqrt{s_{NN}}=13$ TeV.
Second, while the general trends of the dependence on various variables are similar in the $pp$ UPC and $ep$ cases, 
the former allows to probe values of $W$ exceeding those achieved 
in the $ep$ case by at least a factor of ten and to
produce dijets with the significantly larger $M_{12}$ and $M_X$.
In addition, the contribution of the low-$z_{\Pomeron}^{\rm jets}$ region
is much more important in the $pp$ UPC case than in the $ep$ case, which signals the enhanced sensitivity to the
proton diffractive PDFs $f^{D(4)}_{b/p}(x_{\Pomeron},z_{\Pomeron},t,\mu^2)$ at small $z_{\Pomeron}$, where they are poorly constrained~\cite{Chekanov:2004hy,Aktas:2006hy,Aktas:2006hx}.

It is interesting to speculate that the large cross section of diffractive dijet photoproduction in $pp$ UPCs at small
$z_{\Pomeron}^{\rm jets}$ might contribute to hard diffractive dijet production in $pp$ scattering at 
large $\log_{10}(x_{\Pomeron}^{-})$ and cause the dependence of the suppression factor parametrizing the
rapidity gap survival probability on the momentum fraction of the parton in the Pomeron~\cite{Klasen:2009bi,Klasen:2010we}.

In the experiment, to trigger on the events corresponding to diffractive dijet photoproduction in UPCs, 
one should employ the selection criteria
typical for diffractive scattering in photoproduction: the absence of hadronic activity adjacent to the beam directions
(presence of large rapidity gaps)
will correspond to 
(very) small transverse momenta of the final-state protons (ions) and  guarantee that the exchanged photon 
is quasi-real and that the photon--target interaction is diffractive; the detector will register two hard jets with large $p_T$ 
and possibly the forward energy flow from the remnant diffractive final state (see Fig.~\ref{fig:photo_dijets}).

\newpage 

\section{Diffractive dijet photoproduction in proton--nucleus UPCs at the LHC}
\label{sec:pA}

\subsection{General expression for the cross section}
\label{subsec:pA_cs}

Taking one of the ions in Fig.~\ref{fig:photo_dijets} to be a nucleus and the other one to be the proton, 
the cross section of diffractive dijet photoproduction in $pA$ UPCs reads (see Eq.~(\ref{eq:pp_inverse})):
\begin{eqnarray}
d \sigma(pA \to p/A+2{\rm jets}+X^{\prime}+Y) &=& d\sigma(pA\to A+2{\rm jets}+X^{\prime}+Y)^{(+)} \nonumber\\
&+& d \sigma(pA\to p+2{\rm jets}+X^{\prime}+Y)^{(-)} \,.
\label{eq:pA_inverse}
\end{eqnarray}
The first term in Eq.~(\ref{eq:pA_inverse}) corresponds to the process, when the photon flux is produced by the nucleus
and the diffractive photoproduction of dijets takes place on the proton: 
\begin{eqnarray}
d \sigma(pA\to A &+&2{\rm jets}+X^{\prime}+Y)^{(+)}
=\sum_{a,b} \int_{t_{\rm cut}}^{t_{\rm min}} dt \int_{x_{\Pomeron}^{\rm min}}^{x_{\Pomeron}^{\rm max}} dx_{\Pomeron}
\int_0^1 dz_{\Pomeron} \int_{y_{\rm min}}^{y_{\rm max}} dy \int_0^1 dx_{\gamma}
\nonumber\\
&\times & f_{\gamma/A}(y)  f_{a/\gamma}(x_{\gamma},\mu^2) f^{D(4)}_{b/p}(x_{\Pomeron},z_{\Pomeron},t,\mu^2)
d \hat{\sigma}_{ab \to {\rm jets}}^{(n)} \,.
\label{eq:Ap_upc}
\end{eqnarray}
Equation~(\ref{eq:Ap_upc}) is obtained from  Eq.~(\ref{eq:pp_upc}) by replacing the photon flux of the proton $f_{\gamma/p}(y)$
by the photon flux of the nucleus $f_{\gamma/A}(y)$ and 
effectively absorbing in it the factor of $S^2(y)$.

The second term in Eq.~(\ref{eq:pA_inverse}) corresponds to the process, when the photon flux is produced by the proton
and the diffractive photoproduction of dijets takes place on the nucleus:
\begin{eqnarray}
d \sigma(pA\to p &+&2{\rm jets}+X^{\prime}+Y)^{(-)}
=\sum_{a,b} \int_{t_{\rm cut}}^{t_{\rm min}} dt \int_{x_{\Pomeron}^{\rm min}}^{x_{\Pomeron}^{\rm max}} dx_{\Pomeron}
\int_0^1 dz_{\Pomeron} \int_{y_{\rm min}}^{y_{\rm max}} dy \int_0^1 dx_{\gamma}
\nonumber\\
&\times &  f_{\gamma/p}(y)  f_{a/\gamma}(x_{\gamma},\mu^2) f^{D(4)}_{b/A}(x_{\Pomeron},z_{\Pomeron},t,\mu^2)
d \hat{\sigma}_{ab \to {\rm jets}}^{(n)} \,,
\label{eq:pA_upc}
\end{eqnarray}
where $f^{D(4)}_{b/A}$ is the diffractive PDF on a nucleus~\cite{Frankfurt:2011cs}; it a novel, yet unmeasured distribution.
Note that the photon flux $f_{\gamma/p}(y)$ in Eq.~(\ref{eq:pA_upc}) corresponds to $pA$ UPCs and includes the effect of the suppression
of the strong photon--nucleus interaction at central impact parameters (see Sec.~\ref{subsec:pA_flux}).

In Eqs.~(\ref{eq:Ap_upc}) and (\ref{eq:pA_upc}), the jet pseudorapidities $\eta_1$ and $\eta_2$ 
are defined with respect to the direction of the diffracting hadron. Therefore, 
$d \sigma(pA\to p +2{\rm jets}+X^{\prime}+Y)^{(-)}$ is obtained from $d \sigma(pA\to A +2{\rm jets}+X^{\prime}+Y)^{(+)}$
by the appropriate replacements of the photon flux and the diffractive PDFs as explained above and the inversion of
the sign of $\eta_1$ and $\eta_2$: $\eta_{1,2} \to -\eta_{1,2}$ (compare to Eq.~(\ref{eq:reverse_rule})).

It is important to point out that for most of the observables or in the case of the integrated cross section, 
the $pA$ UPC cross section Eq.~(\ref{eq:pA_inverse}) is dominated by the 
first term $d \sigma(pA \to A+2{\rm jets}+X^{\prime}+Y)^{(+)}$. Indeed,  while the photon flux of a nucleus scales as $Z^2$, 
where $Z$ is the nucleus charge, see, e.g.~Eq.~(\ref{eq:flux_proton_PL}), the diffractive PDFs of a nucleus after integration 
over the momentum transfer $t$ scale only as
$A^{4/3}$ in the impulse approximation; they are also further suppressed by nuclear shadowing~\cite{Frankfurt:2011cs}.
Therefore, $pA$ UPCs can be used to primarily study diffractive photoproduction of dijets on the proton by taking advantage
of the dramatically enhanced intensity of the photon flux compared to $pp$ UPCs.   
The same situation arises in exclusive photoproduction of $J/\psi$ mesons in $pA$ UPCs at the LHC at 
$\sqrt{s_{NN}}=5.02$ TeV~\cite{TheALICE:2014dwa}.

\subsection{Flux of equivalent photons in $pA$ UPCs}
\label{subsec:pA_flux}

To calculate the photon flux in $pA$ UPCs, one needs to take into account the suppression of the strong interaction 
between the colliding proton and the nucleus. The resulting photon flux of the 
ultrarelativistic nucleus reads, see e.g.~\cite{Guzey:2013taa}:
\begin{eqnarray}
f_{\gamma/A}(x) &=& \int d^2 b \, \Gamma_{pA}(b) f_{\gamma/A}(x,b) \,,
\label{eq:flux_Ap}
\end{eqnarray} 
where $b$ is the impact parameter (the transverse distance between the centers of mass of the nucleus and the proton);
$\Gamma_{pA}(b)$ is the probability to not have the strong $pA$ interaction at the impact parameter $b$; and
 $f_{\gamma/A}(x,b)$ is the impact parameter dependent photon flux of the nucleus.
 The probability $\Gamma_{pA}(b)$ is given by the standard expression of the Glauber model for high-energy proton--nucleus scattering~\cite{Glauber:1970jm}:
 \begin{equation}
 \Gamma_{pA}(b)=e^{-\sigma_{NN}^{\rm tot}(s) T_A(b)} \,,
 \label{eq:Gamma_pA}
 \end{equation}
where $\sigma_{NN}^{\rm tot}(s)$ is the total nucleon--nucleon cross section; $T_A(b)=\int dz \rho_A(b,z)$ is the nuclear optical 
density, where $\rho_A$ is the density of nucleons; and $\int d^2 b \,T_A(b)=A$, where $A$ is the atomic mass number.
In our analysis, we use the fit of Ref.~\cite{Agashe:2014kda}  for $\sigma_{NN}^{\rm tot}(s)$,
the two-parameter Fermi model parametrization for $\rho_A(b,z)$~\cite{DeJager:1987qc}, and Eq.~(\ref{eq:flux_vidovic}) for 
$f_{\gamma/A}(x,b)$.

\begin{figure}[ht]
\begin{center}
\epsfig{file=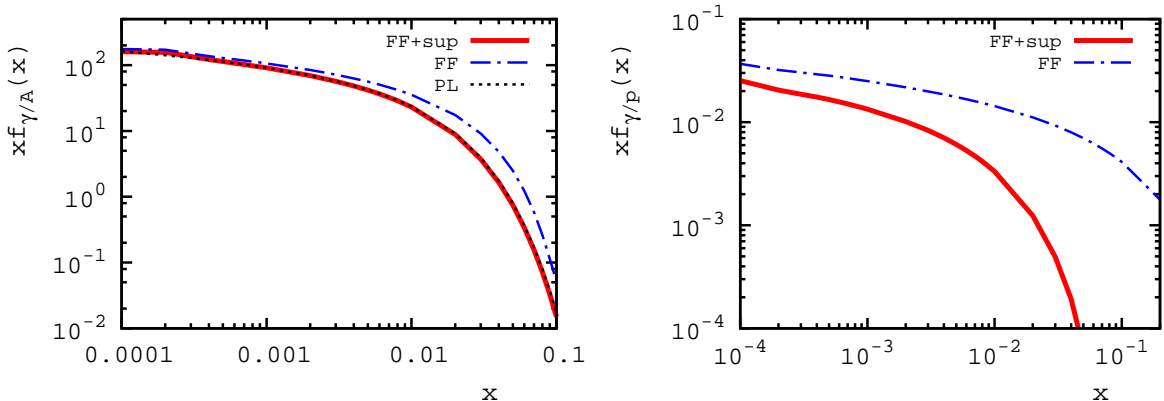,scale=1.25}
 \caption{Left: The photon spectrum $x f_{\gamma/A}(x)$ of Pb in $pA$ UPCs at 
$\sqrt{s_{NN}}=5.02$ TeV as a function of the photon momentum fraction $x$.
 The red solid curve (``FF+sup") is calculated using Eqs.~(\ref{eq:flux_Ap}) and (\ref{eq:Gamma_pA}); the blue dot-dashed
curve (``FF") corresponds to setting $\Gamma_{pA}(b)=1$ in Eq.~(\ref{eq:flux_Ap}); the dotted curve (``PL") is the flux
of a point-like charge, Eq.~(\ref{eq:flux_proton_PL}), with $b_{\rm min}=1.15 R_A$.
Right:
 The photon spectrum $x f_{\gamma/p}(x)$ of the proton in $pA$ UPCs.
For the labels, see the left panel and text.
}
 \label{fig:flux_pA_Dec2015}
\end{center}
\end{figure}

Figure~\ref{fig:flux_pA_Dec2015} (left) shows the photon flux of Pb in $pA$ UPCs at 
$\sqrt{s_{NN}}=5.02$ TeV
as a function of the photon momentum fraction $x$.
The red solid curve labeled ``FF+sup" corresponds to Eqs.~(\ref{eq:flux_Ap}) and (\ref{eq:Gamma_pA}); the blue dot-dashed
curve labeled ``FF" corresponds to the calculation when one sets $\Gamma_{pA}(b)=1$.

In practice, the result of the full calculation can be approximated very well (with a few percent accuracy)
by the photon flux of a point-like charge Eq.~(\ref{eq:flux_proton_PL}) with $b_{\rm min}=1.15 R_A$, 
where $R_A=1.145 A^{1/3} \approx 6.8$ fm is the equivalent sharp radius of $^{208}$Pb.
The corresponding flux is given by the dotted curve labeled ``PL", which is indistinguishable from the red solid curve.
Note that since $\Gamma_{pA}(b)$ is a slow function of $s_{NN}$ in the considered energy range, it is a good approximation to
use $b_{\rm min}=1.15 R_A$ both at $\sqrt{s_{NN}}=5.02$ TeV and $\sqrt{s_{NN}}=8.16$ TeV,
the tentative Run-2 energy of $pA$ collisions at the LHC.

The photon flux of the proton in $pA$ UPCs can be found using Eq.~(\ref{eq:flux_Ap}), where one replaces
$f_{\gamma/A}(x,b)$ by $f_{\gamma/p}(x,b)$. The result is shown by the red solid curve labeled ``FF+sup"
in the right panel of Fig.~\ref{fig:flux_pA_Dec2015}. For comparison, the blue dot-dashed curve labeled ``FF" shows
the result, when one sets $\Gamma_{pA}=1$. Note that this curve coincides with the ``DZ" and ``PL" curves of 
Fig.~\ref{fig:photon_flux_Dec2015} to a few percent accuracy. For convenience, in the Appendix we give a simple 
analytic form of the factor of $f_p^{\rm sup}(x)$ parametrizing
 the difference between the ``FF+sup" and ``FF" curves shown in the right panel of 
Fig.~\ref{fig:flux_pA_Dec2015}. Note that $f_p^{\rm sup}(x)$ does not change when one increases $\sqrt{s_{NN}}$
from $\sqrt{s_{NN}}=5.02$ TeV to $\sqrt{s_{NN}}=8.16$ TeV to better than a fraction of a percent accuracy.

One can readily see from  Fig.~\ref{fig:flux_pA_Dec2015} that $f_{\gamma/A}(x)$ is larger than $f_{\gamma/p}(x)$
by approximately a factor of $5,000$ due to the $Z^2$ factor in Eq.~(\ref{eq:flux_vidovic}). This enhancement 
of the $d \sigma(pA\to A +2{\rm jets}+X^{\prime}+Y)^{(+)}$ term in Eq.~(\ref{eq:Ap_upc}) wins over 
the nuclear enhancement of nuclear diffractive PDFs (see below) entering the $d \sigma(pA\to p +2{\rm jets}+X^{\prime}+Y)^{(-)}$ 
term. As a result, the process, when the photon flux is produced by the nucleus, dominates the cross section of 
$pA$ UPCs unless one probes very large values of $W$ and $\langle \eta^{\rm jets} \rangle$ (see the results
and discussion in Sec.~\ref{subsec:pA_results}).

\subsection{Nuclear diffractive PDFs}
\label{subsec:ndpdfs}

The nuclear diffractive PDFs $f^{D(4)}_{b/A}(x_{\Pomeron},z_{\Pomeron},t,\mu^2)$ entering the $d \sigma(pA\to p +2{\rm jets}+X^{\prime}+Y)^{(-)}$ term in Eq.~(\ref{eq:Ap_upc}) are conditional leading twist PDFs giving the distribution of a parton
$b$ in a nucleus in terms of the light-cone momentum fraction $z_{\Pomeron}$ at the resolution scale $\mu$, provided 
that the nucleus undergoes diffractive scattering characterized by the light-cone momentum fraction loss $x_{\Pomeron}$ and
the invariant momentum transfer squared $t$.

In the impulse approximation (IA), when the only nuclear effect is nuclear coherence, 
$f^{D(4)}_{b/A}(x_{\Pomeron},z_{\Pomeron},t,\mu^2)$ reads:
\begin{equation}
f^{D(4), {\rm IA}}_{b/A}(x_{\Pomeron},z_{\Pomeron},t,\mu^2)=A^2 F_A^2(t) f^{D(4)}_{b/p}(x_{\Pomeron},z_{\Pomeron},t_{\rm min},\mu^2) \,,
\label{eq:ndpfs}
\end{equation}
 where $F_A(t)$ is the nuclear form factor and $t_{\min}=-(x_{\Pomeron} m_p)^2/(1-x_{\Pomeron})$ is the minimal momentum 
 transfer.

At high energies, the hard probe interacts coherently (simultaneously) with all nucleons of a nuclear target, which
results in the effect of nuclear shadowing reducing nuclear PDFs compared to their IA expressions. In particular,
the model of leading twist nuclear shadowing predicts a very significant suppression of nuclear
diffractive PDFs~\cite{Frankfurt:2011cs}, which can be quantified by the suppression factor of $R_b(x_{\Pomeron},z_{\Pomeron},\mu^2)$:
\begin{equation}
f^{D(4)}_{b/A}(x_{\Pomeron},z_{\Pomeron},t,\mu^2)=R_b(x_{\Pomeron},z_{\Pomeron},\mu^2) f^{D(4), {\rm IA}}_{b/A}(x_{\Pomeron},z_{\Pomeron},t,\mu^2)  \,.
\label{eq:ndpfs_2}
\end{equation}
Note that nuclear shadowing of nuclear diffractive PDFs breaks the phenomenological factorization Eq.~(\ref{eq:Regge_fact}) of diffractive PDFs into
the product of the ``Pomeron" flux and PDFs of the ``Pomeron".

\begin{figure}[ht]
\begin{center}
\epsfig{file=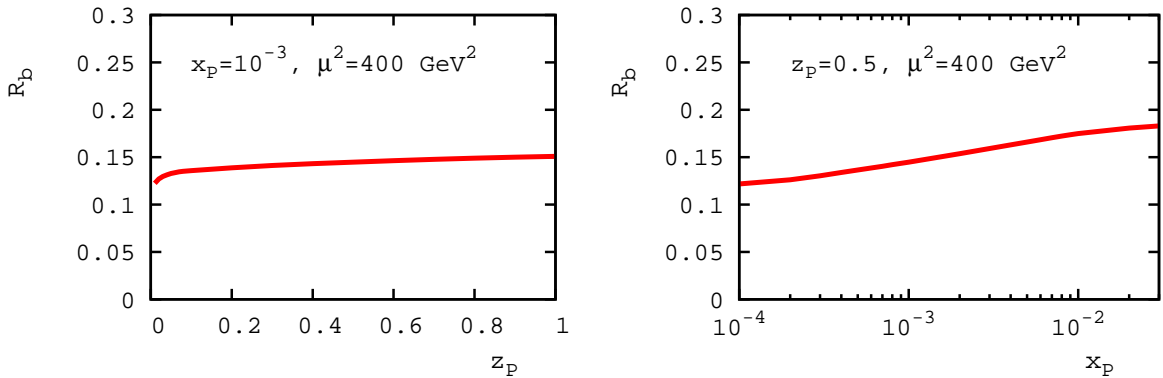,scale=1.25}
 \caption{The suppression factor of $R_b(x_{\Pomeron},z_{\Pomeron},\mu^2)$, Eq.~(\ref{eq:ndpfs_2}), quantifying
 the effect of nuclear shadowing of 
 nuclear diffractive PDFs.}
 \label{fig:Rb}
\end{center}
\end{figure}

Predictions for $R_b(x_{\Pomeron},z_{\Pomeron},\mu^2)$~\cite{Frankfurt:2011cs} for sea quarks for the representative 
ranges of $z_{\Pomeron}$ and $x_{\Pomeron}$ and at $\mu^2=400$ GeV$^2$
are shown in Fig.~\ref{fig:Rb}. An analysis reveals that $R_b(x_{\Pomeron},z_{\Pomeron},\mu^2)$ very weakly depends on 
the parton flavor $b$, the scale $\mu$, $z_{\Pomeron}$ and $x_{\Pomeron}$ (the latter is seen from Fig.~\ref{fig:Rb}).
Therefore, in practical estimates, it is a good approximation to use the constant suppression factor: 
\begin{equation}
R_b(x_{\Pomeron},z_{\Pomeron},\mu^2) \approx 0.15 \,.
\label{eq:ndpfs_3}
\end{equation}

\subsection{Results}
\label{subsec:pA_results}

\begin{figure}[ht]
\begin{center}
\epsfig{file=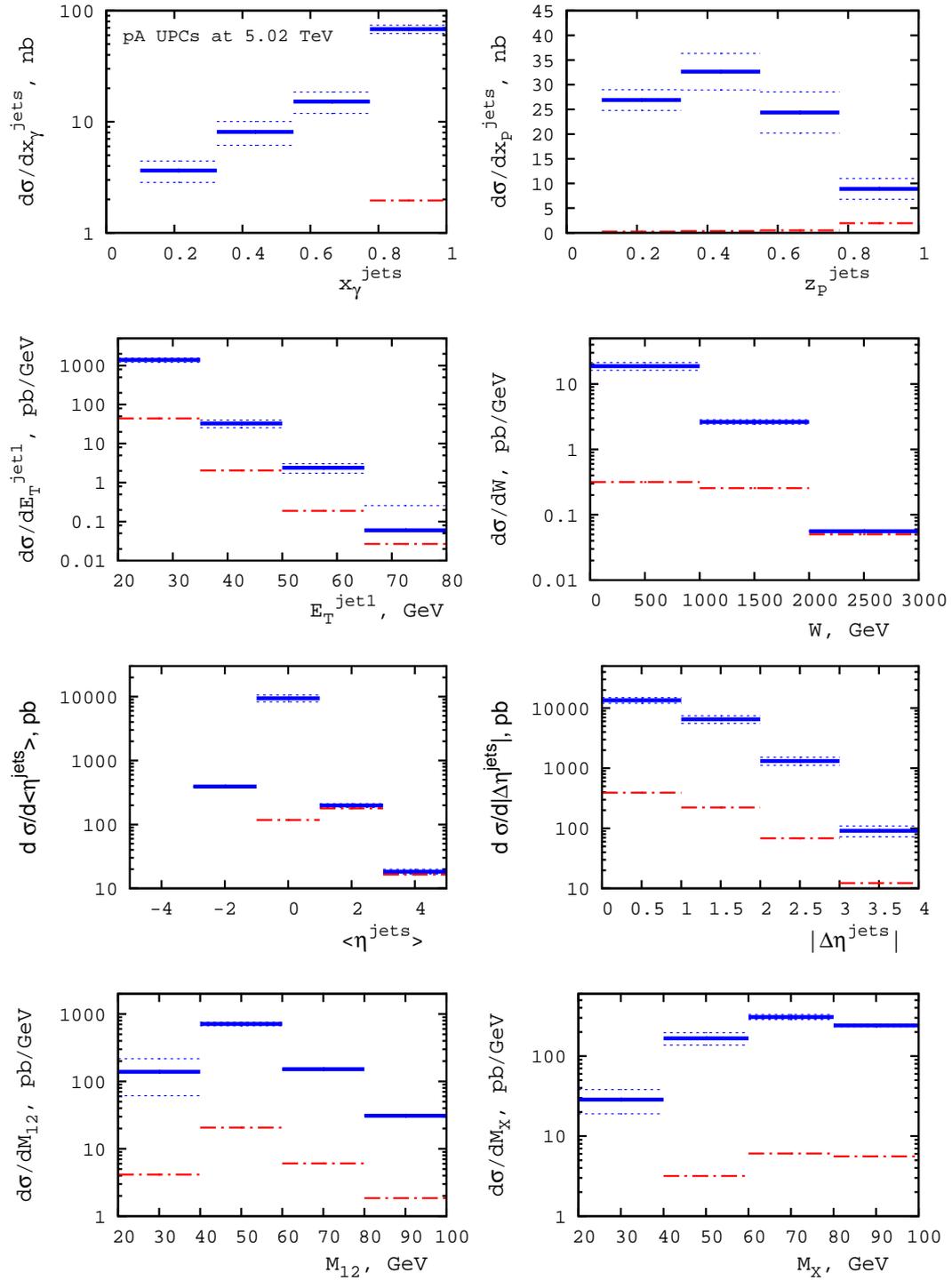,scale=1.1}
 \caption{The differential cross section of diffractive photoproduction of dijets $d \sigma(pA\to p/A+2{\rm jets}+ X^{\prime}+Y)$ 
 in $pA$ UPCs at $\sqrt{s_{NN}}=5.02$ TeV. 
 The net result of Eq.~(\ref{eq:pA_inverse}) (blue solid and dotted lines) and the photon-from-proton contribution only
 (red dot-dashed lines) are shown separately.
 }
 \label{fig:upc_pA_scale}
\end{center}
\end{figure}

\begin{figure}[ht]
\begin{center}
\epsfig{file=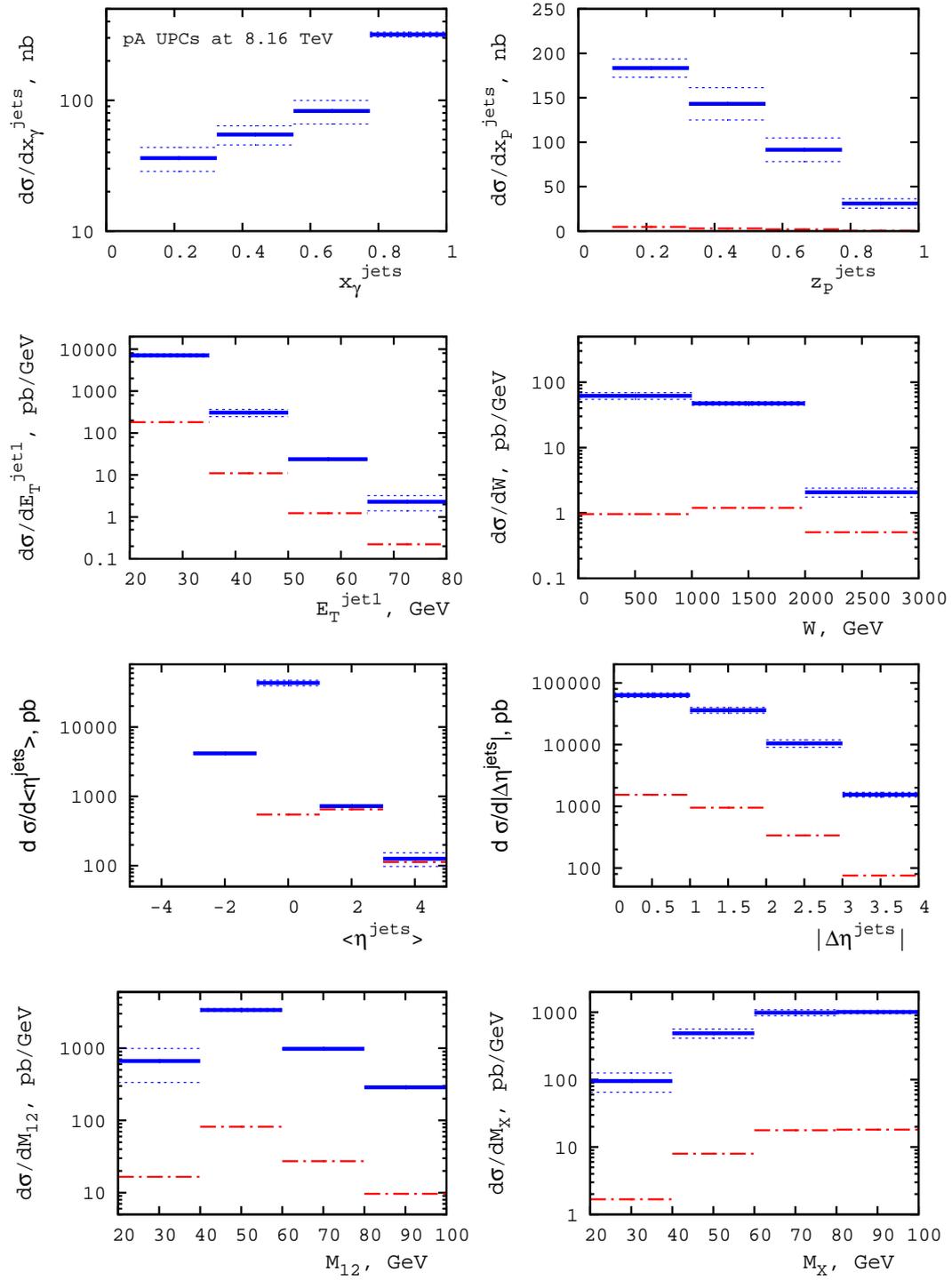,scale=1.1}
 \caption{The same as in Fig.~\ref{fig:upc_pA_scale}, but at $\sqrt{s_{NN}}=8.16$ TeV.}
 \label{fig:upc_pA88_scale}
\end{center}
\end{figure}

We performed NLO calculations of the cross section of diffractive photoproduction of dijets in $pA$ UPCs 
at $\sqrt{s_{NN}}=5.02$ TeV and $\sqrt{s_{NN}}=8.16$ TeV using 
the cuts of Eq.~(\ref{eq:UPC_lj}). The results are presented in Figs.~\ref{fig:upc_pA_scale} 
and \ref{fig:upc_pA88_scale}, respectively.

In these figures, the blue solid and dotted lines give the net result of Eq.~(\ref{eq:pA_inverse}); the red dot-dashed lines show the contribution
of the second term $d \sigma(pA\to p +2{\rm jets}+X^{\prime}+Y)^{(-)}$ in Eq.~(\ref{eq:pA_inverse}) corresponding
to the photon flux emitted by the proton.
The blue solid and red dot-dashed lines correspond to $\mu=E_{T}^{\rm jet 1}$; the two blue dotted lines surrounding each
solid line correspond to $\mu=2 E_{T}^{\rm jet 1}$ and $\mu=E_{T}^{\rm jet 1}/2$, respectively, which 
demonstrates the theoretical uncertainty of our predictions associated with the choice of the renormalization and factorization scale.
One can see from the figures that this uncertainty is not significant.

A comparison of our predictions for $pA$ UPCs shown in Figs.~\ref{fig:upc_pA_scale} and \ref{fig:upc_pA88_scale} to those
for $pp$ UPCs shown in Figs.~\ref{fig:upc_pp_scale_S2} and \ref{fig:upc_pp_scale_13_S2} demonstrates that 
the general trends for the dependence of the cross section on various variables are similar in the $pp$ and $pA$ cases:
very roughly, the $pA$ results can be obtained from the $pp$ ones by multiplying them by the scaling factor of
$(1/2) Z^2 (0.7\ {\rm fm}/R_A) \approx 350$, where took into account that the photon spectra of the proton and 
a nucleus extend up to $x \sim 1/b_{\min}=1/(0.7\ {\rm fm})$ and  $b_{\min} \sim 1/R_A$, respectively, and that 
in the $pA$ UPC cross section, the contribution of the minus-term in Eq.~(\ref{eq:pA_inverse}) is generally small.
Note that in the two upper panels of  Figs.~\ref{fig:upc_pA_scale} 
and \ref{fig:upc_pA88_scale}, the physical units along the $y$-axis are nb.

At the same time, there are marked differences between the $pA$ and $pp$ results. 
First, the cross section falls off faster as one increases $E_{T}^{\rm jet 1}$ or $W$ in the 
$pA$ case than in the $pp$ case because the photon flux of a heavy ultrarelativistic nucleus decreases with an increase 
of the photon energy much faster than the photon flux of the proton. This also explains why at large 
values of $E_{T}^{\rm jet 1}$ and $W$, the cross section is dominated by the photon-from-proton contribution
(the term $d \sigma(pA\to p +2{\rm jets}+X^{\prime}+Y)^{(-)}$ in Eq.~(\ref{eq:pA_inverse}) corresponding
to the photon flux emitted by the proton).

Second, unlike the symmetric $\langle \eta^{\rm jets} \rangle$ distribution in $pp$ UPCs,  this distribution
is not symmetric in the case of $pA$ UPCs. Indeed, at central and backward dijet rapidities, the dominant contribution to
the cross section in Eq.~(\ref{eq:pA_inverse}) comes from low-energy photons emitted by the nucleus. 
At the same time, forward dijet rapidities correspond to high-energy photons emitted by the nucleus, where the photon
flux is very small, or to low-energy photons emitted by the proton; the latter contribution dominates the
cross section for sufficiently large positive $\langle \eta^{\rm jets} \rangle$.

Note that without the effect of nuclear shadowing in nuclear diffractive PDFs, 
the photon-from-proton contribution (the red dot-dashed curves in Figs.~\ref{fig:upc_pA_scale}
and \ref{fig:upc_pA88_scale}) would be globally larger by a factor of $1/0.15 \approx 7$ 
(cf.~Eq.~(\ref{eq:ndpfs_3})), with a weak dependence on $z_{\Pomeron}$ and $x_{\Pomeron}$
     (cf.~Fig.~\ref{fig:Rb}).

\newpage

\section{Diffractive dijet photoproduction in nucleus--nucleus UPCs at the LHC}
\label{sec:AA}

\subsection{General expression for the cross section}

Considering nucleus--nucleus UPCs, we can readily write down the cross section of coherent diffractive dijet photoproduction (compare to
Eqs.~(\ref{eq:pp_inverse}) and (\ref{eq:pA_inverse})):
\begin{eqnarray}
d \sigma(AA \to A+2{\rm jets}+X^{\prime}+A) &=& d\sigma(AA\to A+2{\rm jets}+X^{\prime}+A)^{(+)} \nonumber\\
&+& d \sigma(AA\to A+2{\rm jets}+X^{\prime}+A)^{(-)} \,.
\label{eq:AA}
\end{eqnarray}
Note that we considered the case of coherent nuclear scattering when both nuclei remain intact after either the photon emission
or hard dijet photoproduction.

By analogy with the $pp$ and $pA$ cases considered above, each term in Eq.~(\ref{eq:AA}) can be written 
as a convolution of the photon flux of an ultrarelativistic nucleus $f_{\gamma/A}(y)$, the PDF of the photon $f_{a/\gamma}$, the nuclear diffractive PDFs 
$f^{D(4)}_{b/A}$, and the hard elementary cross section $d \hat{\sigma}_{ab \to {\rm jets}}^{(n)}$,
see Eqs.~(\ref{eq:pp_upc}), (\ref{eq:Ap_upc}) and (\ref{eq:pA_upc}). For instance,  
one obtains for the first term:
\begin{eqnarray}
d \sigma(AA\to A &+&2{\rm jets}+X^{\prime}+A)^{(+)}
=\sum_{a,b} \int_{t_{\rm cut}}^{t_{\rm min}} dt \int_{x_{\Pomeron}^{\rm min}}^{x_{\Pomeron}^{\rm max}} dx_{\Pomeron}
\int_0^1 dz_{\Pomeron} \int_{y_{\rm min}}^{y_{\rm max}} dy \int_0^1 dx_{\gamma}
\nonumber\\
&\times & f_{\gamma/A}(y)  f_{a/\gamma}(x_{\gamma},\mu^2) f^{D(4)}_{b/A}(x_{\Pomeron},z_{\Pomeron},t,\mu^2)
d \hat{\sigma}_{ab \to {\rm jets}}^{(n)} \,.
\label{eq:AA_2}
\end{eqnarray} 
For symmetric (equal beam-energy) $AA$ UPCs that we consider in our work, the second and first terms in Eq.~(\ref{eq:AA}) are related by a sign exchange
of the dijet rapidities:
\begin{equation}
d \sigma(AA\to A+2{\rm jets}+ X^{\prime}+A)^{(-)}=d \sigma(AA\to A +2{\rm jets}+ X^{\prime}+A)_{|\eta_1 \to -\eta_1,\, \eta_2 \to -\eta_2}^{(+)} \,.
\label{eq:reverse_rule_AA}
\end{equation}

\subsection{Flux of equivalent photons in $AA$ UPCs}
\label{subsec:AA_flux}

In the calculation of the photon flux produced by each ultrarelativistic nucleus in $AA$ UPCs, one needs to suppress
the strong nucleus--nucleus interaction. The resulting expression for the photon flux $f_{\gamma/A}(x)$ is
\begin{equation}
f_{\gamma/A}(x)=\int d^2 b \,\Gamma_{AA}(b) f_{\gamma/A}(x,b) \,,
\label{eq:flux_AA}
\end{equation}
where $f_{\gamma/A}(x,b)$ is the impact parameter dependent photon flux of the nucleus~(\ref{eq:flux_vidovic}) and
$\Gamma_{AA}(b)$ is the probability for the nuclei to not interact strongly at the impact parameter $b$.
It is given by the standard expression of the Glauber model for high-energy nucleus--nucleus scattering:
\begin{equation}
\Gamma_{AA}(b)=\exp\left(-\sigma_{NN}^{\rm tot}(s) \int d^2 \vec{b}_1 T_A(\vec{b}) T_A(\vec{b}_1-\vec{b})\right) \,.
\label{eq:Gamma_AA}
\end{equation}

\begin{figure}[t]
\begin{center}
\epsfig{file=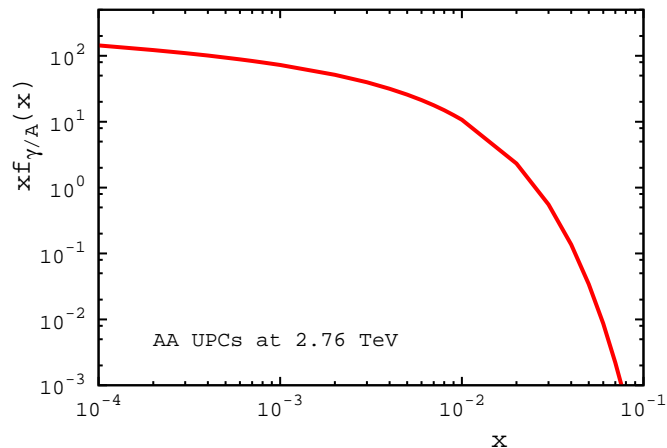,scale=1.0}
 \caption{The photon spectrum $x f_{\gamma/A}(x)$ of Pb in $AA$ UPCs at $\sqrt{s_{NN}}=2.76$ TeV as a function
 of the photon momentum fraction $x$.}
 \label{fig:photon_flux_AA}
\end{center}
\end{figure}

An analysis shows that the result of the calculation of the photon flux using the exact expression of Eq.~(\ref{eq:flux_AA}) can be very well approximated by the much simpler expression for the photon flux produced by a relativistic point-like charge $Z$ in Eq.~(\ref{eq:flux_proton_PL}), when one uses $b_{\rm min} \approx 2 R_A$ for the minimal impact parameter,
where $R_A$ is the equivalent sharp nucleus radius. Therefore, in our analysis of Pb-Pb UPCs at the LHC,
we used Eq.~(\ref{eq:flux_proton_PL}) for the Pb photon flux in Eq.~(\ref{eq:AA_2}) with $b_{\rm min} \approx 2.1 R_A=14.2$ fm~\cite{Nystrand:2004vn}.  
The resulting photon spectrum $x f_{\gamma/A}(x)$ as a function of the energy fraction $x$ is shown in Fig.~\ref{fig:photon_flux_AA}. Note that in our analysis we assume that $f_{\gamma/A}(x)$ does not change when one increases the invariant collision energy from $\sqrt{s_{NN}}=2.76$ TeV to $\sqrt{s_{NN}}=5.1$ TeV, the tentative Run-2 energy of 
$AA$ collisions at the LHC.

\subsection{Results}
\label{subsec:AA_results}

\begin{figure}[ht]
\begin{center}
\epsfig{file=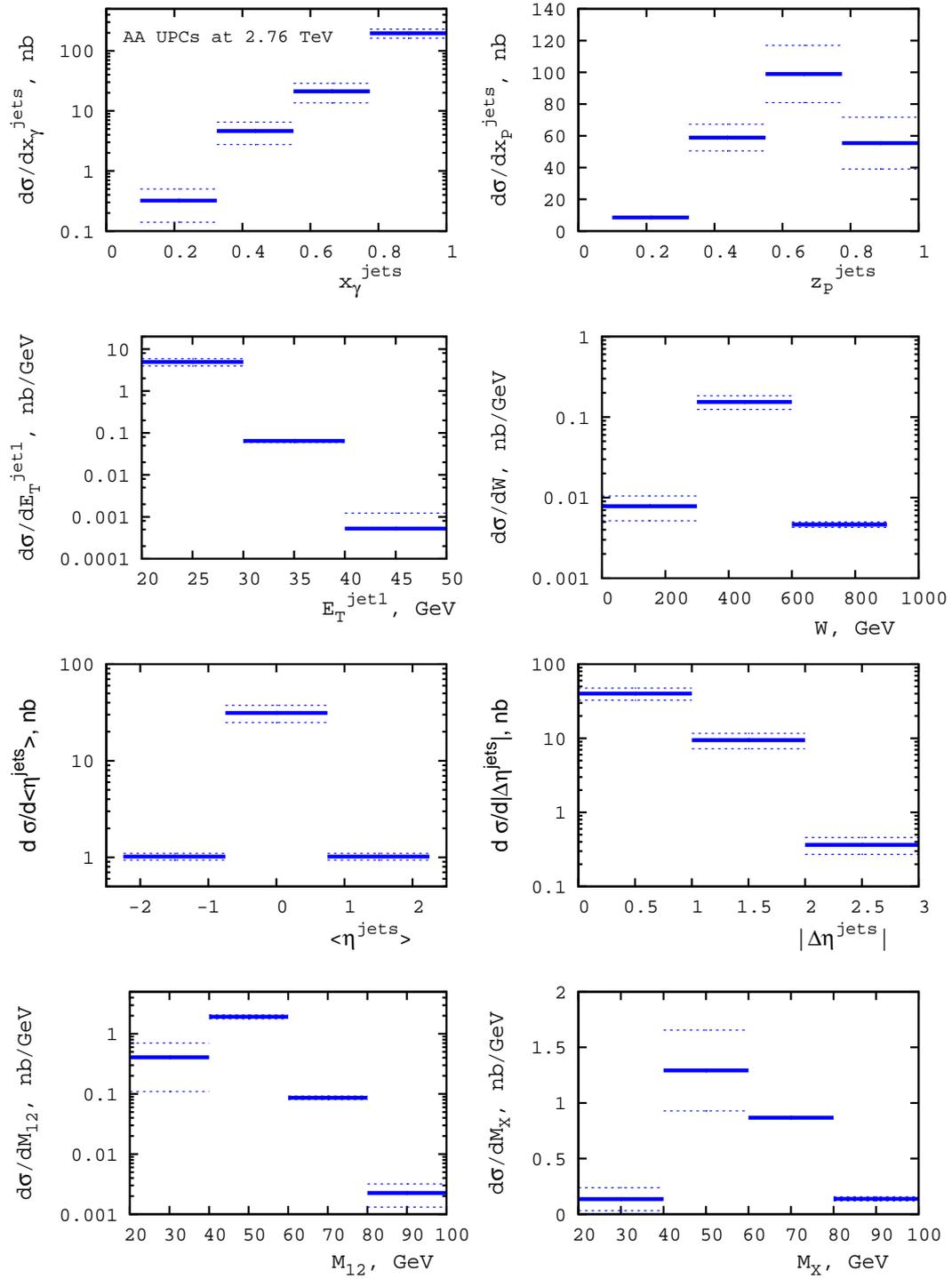,scale=1.1}
 \caption{The differential cross section of diffractive photoproduction of dijets $d \sigma(AA\to A+2{\rm jets}+ X^{\prime}+A)$
 in $AA$ UPCs at $\sqrt{s_{NN}}=2.76$ TeV.}
 \label{fig:upc_AA}
\end{center}
\end{figure}

\begin{figure}[ht]
\begin{center}
\epsfig{file=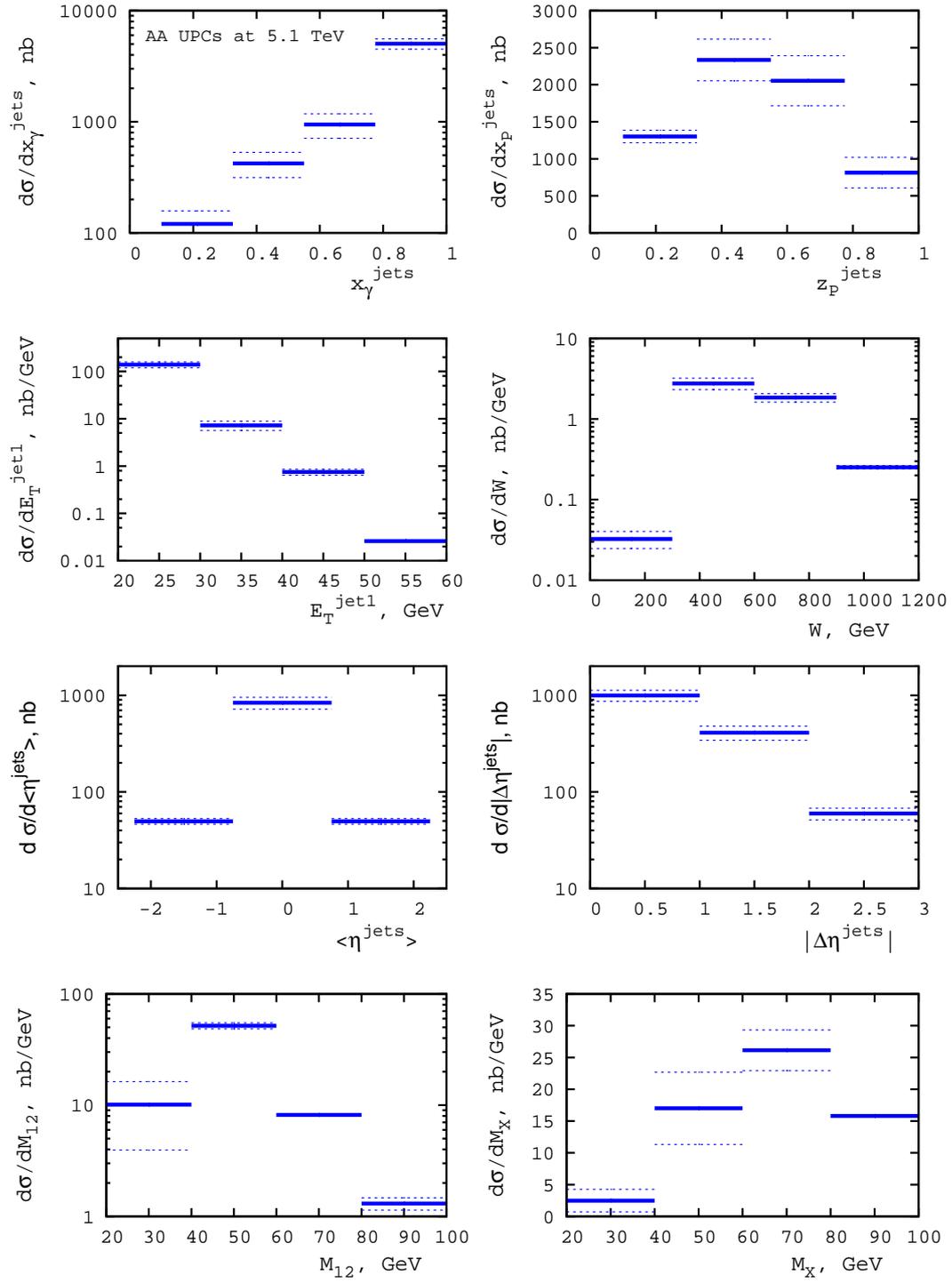,scale=1.1}
 \caption{The same as Fig.~\ref{fig:upc_AA}, but at $\sqrt{s_{NN}}=5.1$ TeV.}
 \label{fig:upc_AA5}
\end{center}
\end{figure}

The results of our NLO calculations of the cross section of diffractive photoproduction of dijets in $AA$ UPCs 
at $\sqrt{s_{NN}}=2.76$ TeV and $\sqrt{s_{NN}}=5.1$ TeV are presented in Figs.~\ref{fig:upc_AA} 
and \ref{fig:upc_AA5}, respectively. In the calculations, use used
the cuts of Eq.~(\ref{eq:UPC_lj}). As in the $pp$ and $pA$ cases, the blue solid lines correspond to the 
$\mu=E_{T}^{\rm jet 1}$ choice of the renormalization and factorization scale; the two dot-dashed lines surrounding the solid one correspond
to $\mu=2 E_{T}^{\rm jet 1}$ and $\mu=E_{T}^{\rm jet 1}/2$ and, thus, illustrate the theoretical uncertainty of our calculations associated with the choice of the two scales.

One can see from these figures that the general trends of the dependence of the cross section of diffractive dijet photoproduction
in $AA$ UPCs resemble closely those already observed in the $pp$ and $pA$ and can be obtained 
approximately
by simple rescaling.
For instance, a comparison of the $AA$ results at
$\sqrt{s_{NN}}=5.1$ TeV with the $pA$ results at $\sqrt{s_{NN}}=8.16$ TeV (corresponding to the same nucleus beam energy),
one finds that the scaling factor between the distributions in the two cases is approximately $A$.
In this estimate, we took into account that in the bulk of the considered kinematics, one has the approximate relation
$f_{j/A}^{D(3)}(x_{\Pomeron},z_{\Pomeron},\mu^2) \approx A/2  f_{j/p}^{D(3)}(x_{\Pomeron},z_{\Pomeron},\mu^2)$
between the nucleus and proton diffractive PDFs, integrated over the momentum transfer $t$,
(see Eqs.~(\ref{eq:ndpfs})--(\ref{eq:ndpfs_3})),
and that the $AA$ UPC cross section receives contributions of both nuclei, while the $pA$ UPC cross section is dominated 
by the photon-from-nucleus contribution. 

Note that the strong nuclear shadowing suppresses nuclear diffractive PDFs by the factor of $0.15$, see
Eq.~(\ref{eq:ndpfs_3}); without this effect, i.e., in the impulse approximation, our results in Figs.~\ref{fig:upc_AA}
and \ref{fig:upc_AA5} would be increased approximately by the factor of seven.

\newpage

\section{Factorization breaking in diffractive dijet photoproduction}
\label{sec:factor_br}

It is well known from studies of diffractive photoproduction of dijets in $ep$ scattering at HERA that 
collinear factorization for this process is broken, i.e., NLO pQCD calculations
overestimate the measured cross sections by almost a factor of two~\cite{Chekanov:2007rh,Aktas:2007hn,Aaron:2010su,Andreev:2015cwa,Klasen:2004qr,Klasen:2004ct,Klasen:2005dq,Klasen:2008ah,Klasen:2010vk}.
However, the pattern of this factorization breaking remains unknown and presents one of the outstanding questions in this
field: the data and the theory can be made consistent by introducing either the global suppression factor of
$R(\rm glob.) \approx 0.5$ or the suppression factor of $R(\rm res.) \approx 0.4$ only for the resolved photon contribution.

In addition, the HERA data on diffractive photoproduction of open charm~\cite{Chekanov:2007pm} are in agreement with NLO
pQCD calculations, which is consistent with diffractive QCD factorization. This agreement can be interpreted as an indication
of absence of factorization breaking for the direct photon contribution and the charm-quark part of the resolved
photon contribution to the dijet photoproduction cross section. Hence, it challenges the global suppression scenario of diffractive factorization breaking.

Factorization breaking in diffractive dijet photoproduction results from soft inelastic photon interactions with the proton
(nucleus),
which populate and thus partially destroy the final-state rapidity gap. Thus, it has exactly the same nature as the
rapidity gap survival probability $S^2$, which we discussed above in relation to $pp$ UPCs, see Eq.~(\ref{eq:S2_pp}).
At high energies, the photon interacts with protons and nuclei by fluctuating into various 
hadronic configurations (components) interacting with the target with different cross sections.
Thus, it is natural to put forward the following physics scenario~\cite{Kaidalov:2009fp}: for the direct photon contribution,
corresponding to weakly-interacting (point-like) fluctuations of the photon, factorization holds; 
for the resolved photon contribution corresponding to large-size photon fluctuations interacting with a typical vector
meson--nucleon cross section, factorization is broken, which leads to the suppression factor of $R(\rm res.) \approx 0.3-0.4$.
Note that beyond the leading order of pQCD, the separations of the direct and resolved contributions
is ambiguous and depends on the factorization scheme and the factorization 
scale~\cite{Klasen:2005dq,Klasen:2002xb}; in the present work, we use the conventions of~\cite{Klasen:2010vk}.  

\begin{figure}[t]
\begin{center}
\epsfig{file=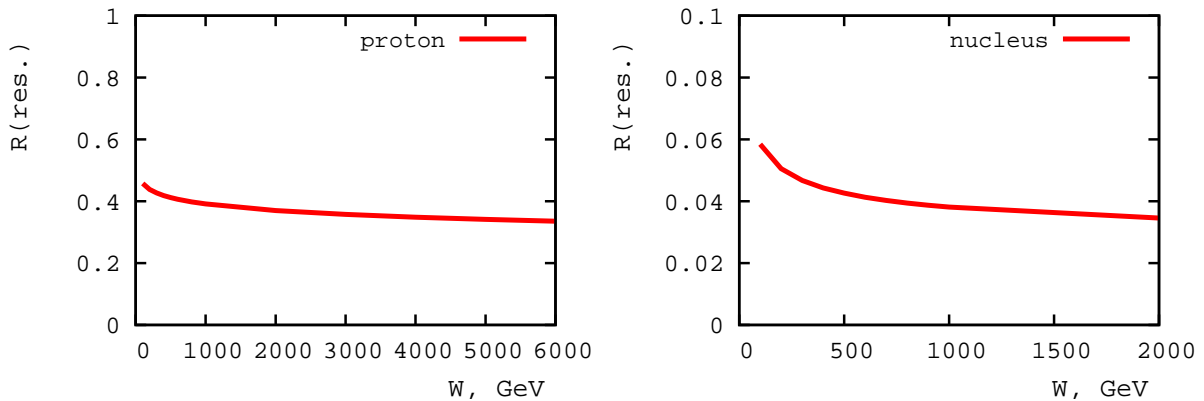,scale=1.25}
 \caption{The factor of $R({\rm res.})$, Eq.~(\ref{eq:R_res}), quantifying the effect of factorization breaking (suppression) for 
 the resolved photon contribution.}
 \label{fig:R_res}
\end{center}
\end{figure}

Our results presented so far in Figs.~\ref{fig:upc_pp_scale_S2}, \ref{fig:upc_pp_scale_13_S2}, \ref{fig:upc_pA_scale},
\ref{fig:upc_pA88_scale}, \ref{fig:upc_AA} and \ref{fig:upc_AA5} assume no factorization breaking.
Based on the observations and arguments summarized above, we will test the following two competing scenarios of diffractive 
QCD factorization breaking and implement them in our predictions for the cross section of diffractive dijet photoproduction:
first, we assume the global suppression factor of $R(\rm glob.) = 0.5$
for the proton target and $R(\rm glob.)=0.1$ for the nucleus target (the latter value is somewhat ad hoc, but 
reflects the important observation that it is much easier to break the nucleus than the proton, 
see Fig.~\ref{fig:R_res} and its discussion below);
second, we assume that
the resolved photon contribution enters with the suppression factor of $R(\rm res.)$, while the direct
photon contribution is unsuppressed. To estimate $R(\rm res.)$, we use the appropriate application of the two-state eikonal 
model of~\cite{Khoze:2000wk,Kaidalov:2003xf}
(compare to Eq.~(\ref{eq:S2_pp})):
\begin{equation}
R({\rm res.})=\frac{\int d^2 b \, |{\cal A}_{\gamma T \to VT}(W,b)|^2 P_{VT}(W,b)}{\int d^2 b \, |{\cal A}_{\gamma T \to VT}(W,b)|^2} \,,
\label{eq:R_res}
\end{equation} 
where $T$ stands for the proton or nucleus target; $V$ denotes the hadron-like fluctuation (component) of the photon, which
is assumed to be represented by the $\rho$ meson; ${\cal A}_{\gamma T \to VT}(W,b)$ is the $\gamma T \to V T$ 
amplitude  in impact parameter space; $P_{VT}(W,b)$ is the probability to not have the strong inelastic vector meson--target
interaction at the impact parameter $b$; and $W$ is the invariant photon--nucleon energy.

For the proton target, we use 
\begin{equation}
|{\cal A}_{\gamma p \to Vp}(W,b)|^2=e^{-b^2/B(W)} |{\cal A}_{\gamma p \to Vp}(W,b=0)|^2 \,,
\label{eq:A_p}
\end{equation}
where $B(W)$ is the slope of the $t$-dependence of the $\gamma p \to \rho p$ cross section. A fit to the available 
HERA data gives $B(W)=[11+0.5 \ln (W/W_0)^2]$ GeV$^{-2}$, where $W_0=72$ GeV~\cite{Aid:1996bs,Breitweg:1997ed}.

For the probability $P_{Vp}(W,b)$, we use Eqs.~(\ref{eq:P}) and (\ref{eq:Omega}), where in the expression for 
the proton optical density, we substitute the total proton--proton cross section $\sigma_{pp}^{\rm tot}(s)$
by the $\rho$ meson--nucleon cross section $\sigma_{\rho N}(W)$. Since we are interested in the 
large values of $W > 100$ GeV well beyond the HERA reach, we use in our analysis the following simple and conservative
extrapolation:
\begin{equation}
\sigma_{\rho N}(W)=26 \left(\frac{W^2}{W_0^2}\right)^{0.08} \ {\rm mb} \,,
\label{eq:sigma_rhop}
\end{equation}
where $W_0=100$ GeV. The value of $\sigma_{\rho N}(W)$ at $W=100$ GeV agrees with the analysis of Ref.~\cite{Frankfurt:2015cwa}.

To find $R({\rm res.})$ for the nuclear target, we calculate ${\cal A}_{\gamma A \to VA}(W,b)$ in Eq.~(\ref{eq:R_res})
using the Glauber model of nuclear shadowing for coherent photoproduction of vector mesons on nuclei in the high-energy limit, see e.g.~\cite{Bauer:1977iq}:
\begin{equation}
{\cal A}_{\gamma A \to VA}(W,b)=\frac{e}{f_{V}}  \left(1-e^{-\frac{\sigma_{\rho N}(W)}{2} T_A(b)}\right) \,,
\label{eq:A_A}
\end{equation}
where $T_A(b)$ is the nuclear optical density normalized to the number of nucleons $A$ and $f_V^2/(4 \pi)=2.01$ is the 
photon--$\rho$ meson coupling constant determined from the $\rho \to e^{+}e^{-}$ decay.
Note that in Eq.~(\ref{eq:A_A}) we neglected the effect of the inelastic (Gribov) nuclear shadowing --- at the large values of 
$W$ that we consider, due to an eventual decrease of the dispersion of hadronic fluctuations of a projectile with 
an increase of energy~\cite{Guzey:2005tk},
the relative importance of inelastic nuclear shadowing in our case is much smaller than that in the case of coherent
$\rho$ and $\phi$ photoproduction in $AA$ UPCs~~\cite{Frankfurt:2015cwa,Guzey:2016piu}. 

For the suppression factor of $P_{VA}(W,b)$ in Eq.~(\ref{eq:R_res}), we use the standard Glauber model expression
for the probability to not have the strong inelastic resolved photon ($\rho$ meson)--nucleus interaction at the impact parameter $b$ (compare to Eq.~(\ref{eq:Gamma_pA})):
\begin{equation}
P_{VA}(W,b)=e^{-\sigma_{\rho N}(W) T_A(b)} \,.
\end{equation}

Figure~\ref{fig:R_res} shows the resulting values of $R({\rm res.})$ for the proton (left panel) and Pb (right panel) as a 
function of the invariant photon--nucleon energy $W$. One can see from the figure that for the proton, 
$R({\rm res.}) \approx 0.4$, which is in agreement with the original result of~\cite{Khoze:2000wk}.
For Pb, the values of $R({\rm res.})$ are an order of magnitude smaller, $R({\rm res.}) \approx 0.04$, which reflects
the very small probability of rapidity gap events with nuclear targets.

While our second scenario involving $R({\rm res.})$ captures the bulk of physics of diffractive factorization
breaking coming from the hadron structure of the photon, it neglects such subtle points as the possible dependence of
$R({\rm res.})$ on the parton flavor and $x_{\gamma}$ due to the separation of the resolved contribution into the point-like and hadronic terms, the hadronization corrections and bin migration effects,  see the discussion in Ref.~\cite{Kaidalov:2009fp}.
Our aim here is to examine whether studies of diffractive dijet photoproduction in UPCs
can help to distinguish between the two scenarios and, thus, to complement and extend the analysis of this process at HERA.

Note that for the first time, the issue of nuclear dependence of factorization breaking in
diffractive dijet production in hard and ultraperipheral $pA$ scattering was considered in~\cite{Guzey:2005ys}. 
It was found that soft inelastic proton--nucleus interactions significantly suppress the rapidity gap probability 
in hard $pA$ scattering, which is in line with the small values of $R(\rm glob.)$ and $R(\rm res.)$ for
the nucleus target, which we use in our analysis.

\begin{figure}[t]
\begin{center}
\epsfig{file=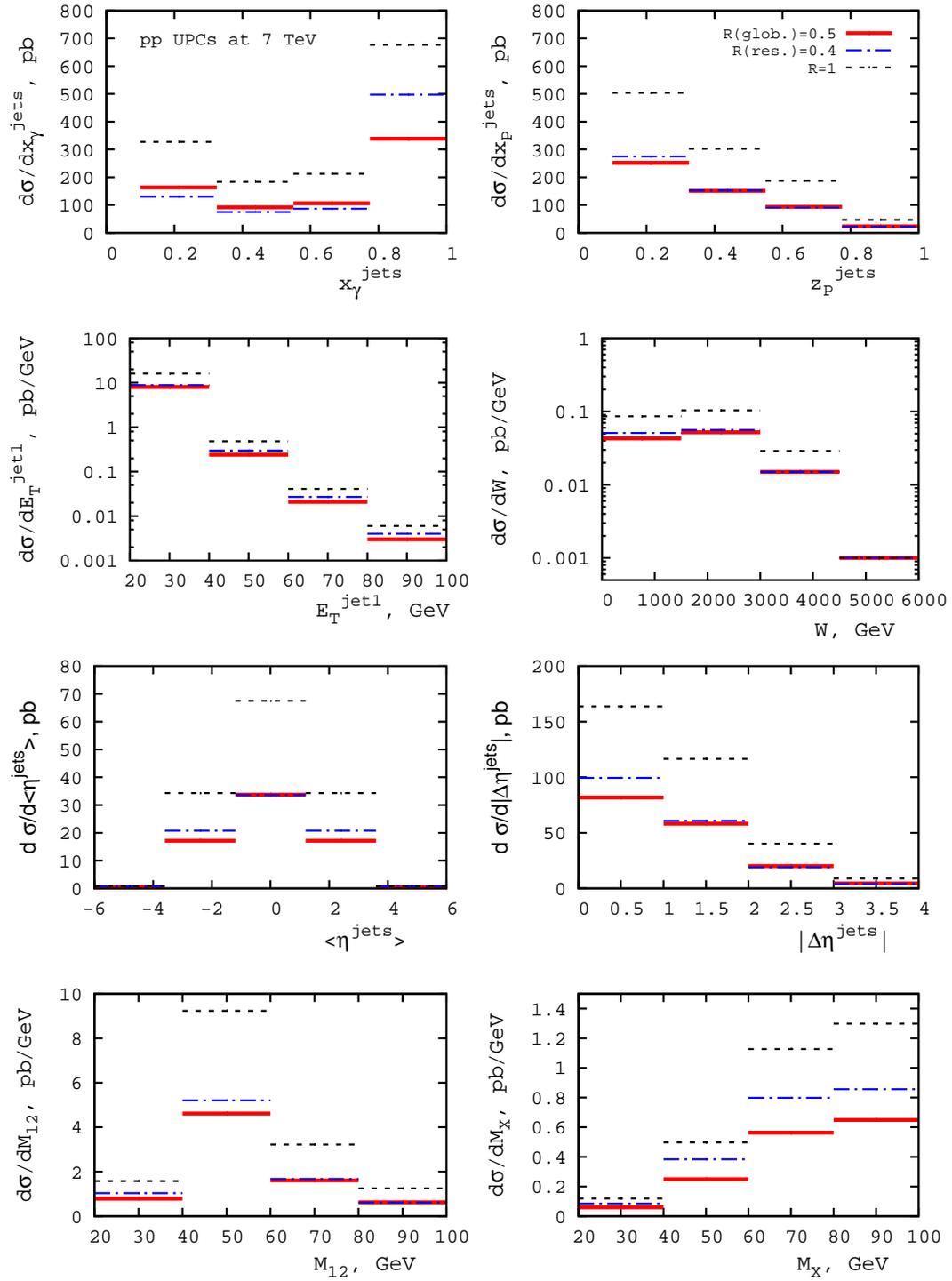,scale=1.1}
 \caption{The effect of diffractive factorization breaking on the differential cross section of 
 diffractive photoproduction of dijets $d \sigma(pp\to p+2{\rm jets}+ X^{\prime}+Y)$
 in $pp$ UPCs at $\sqrt{s_{NN}}=7$ TeV.}
 \label{fig:pp_fb}
\end{center}
\end{figure}

\begin{figure}[t]
\begin{center}
\epsfig{file=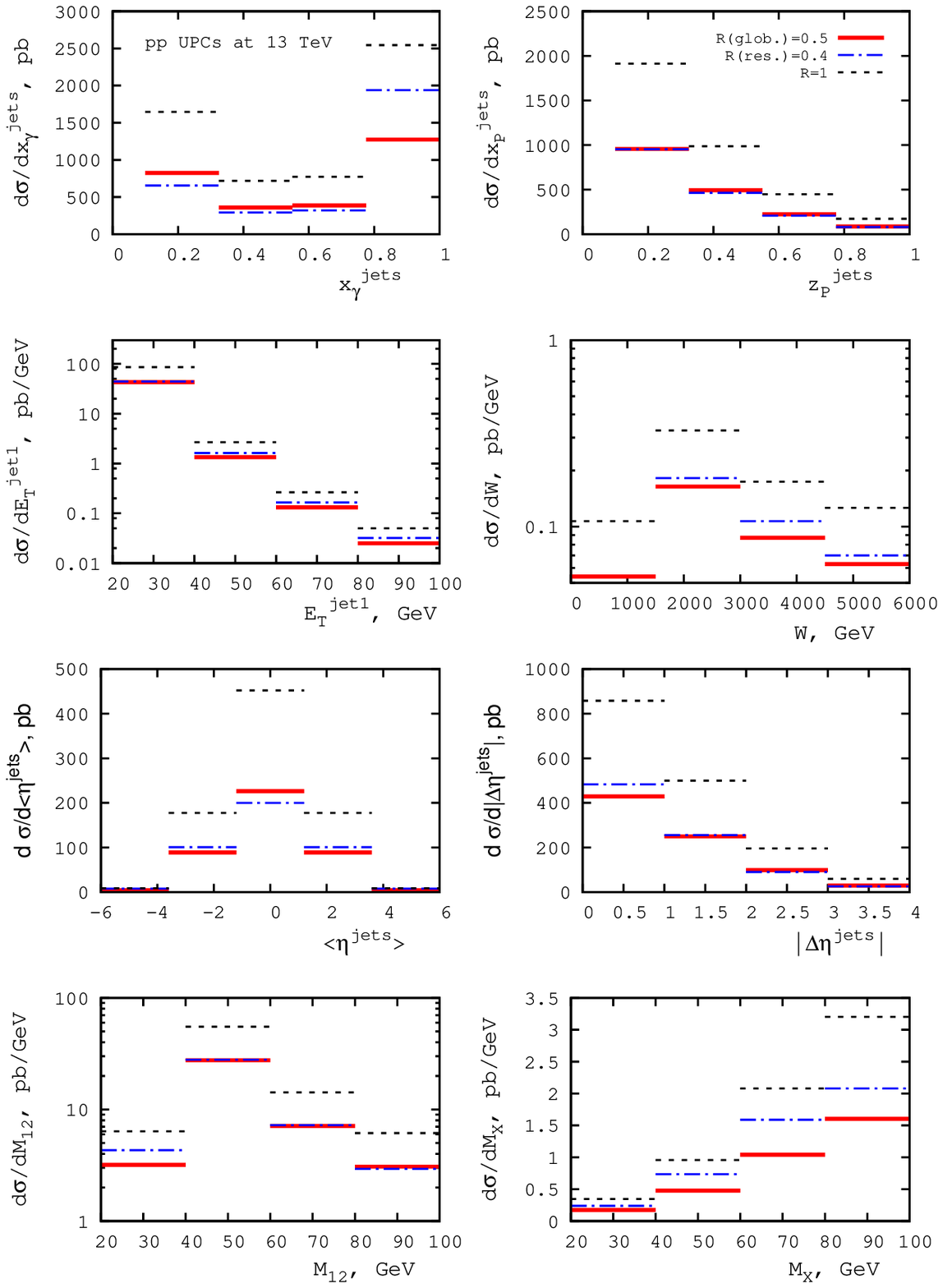,scale=1.1}
 \caption{The same as in Fig.~\ref{fig:pp_fb}, but at $\sqrt{s_{NN}}=13$ TeV.}
 \label{fig:pp13_fb}
\end{center}
\end{figure}

\begin{figure}[ht]
\begin{center}
\epsfig{file=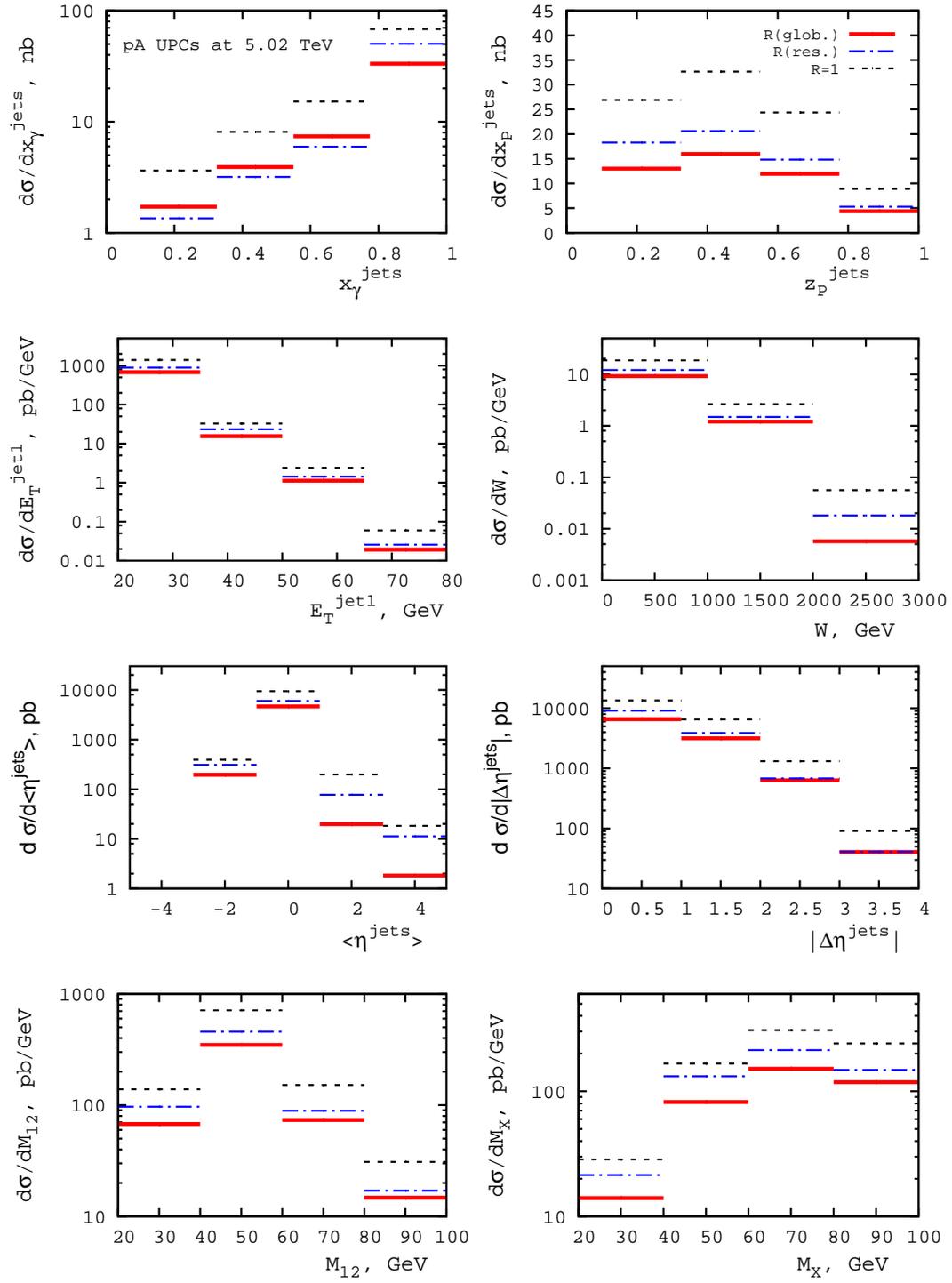,scale=1.1}
 \caption{The effect of diffractive factorization breaking on the cross section of diffractive photoproduction of dijets $d \sigma(pA\to p/A+2{\rm jets}+ X^{\prime}+Y)$ 
 in $pA$ UPCs at $\sqrt{s_{NN}}=5.02$ TeV. 
 }
 \label{fig:pA_fb}
\end{center}
\end{figure}

\begin{figure}[t]
\begin{center}
\epsfig{file=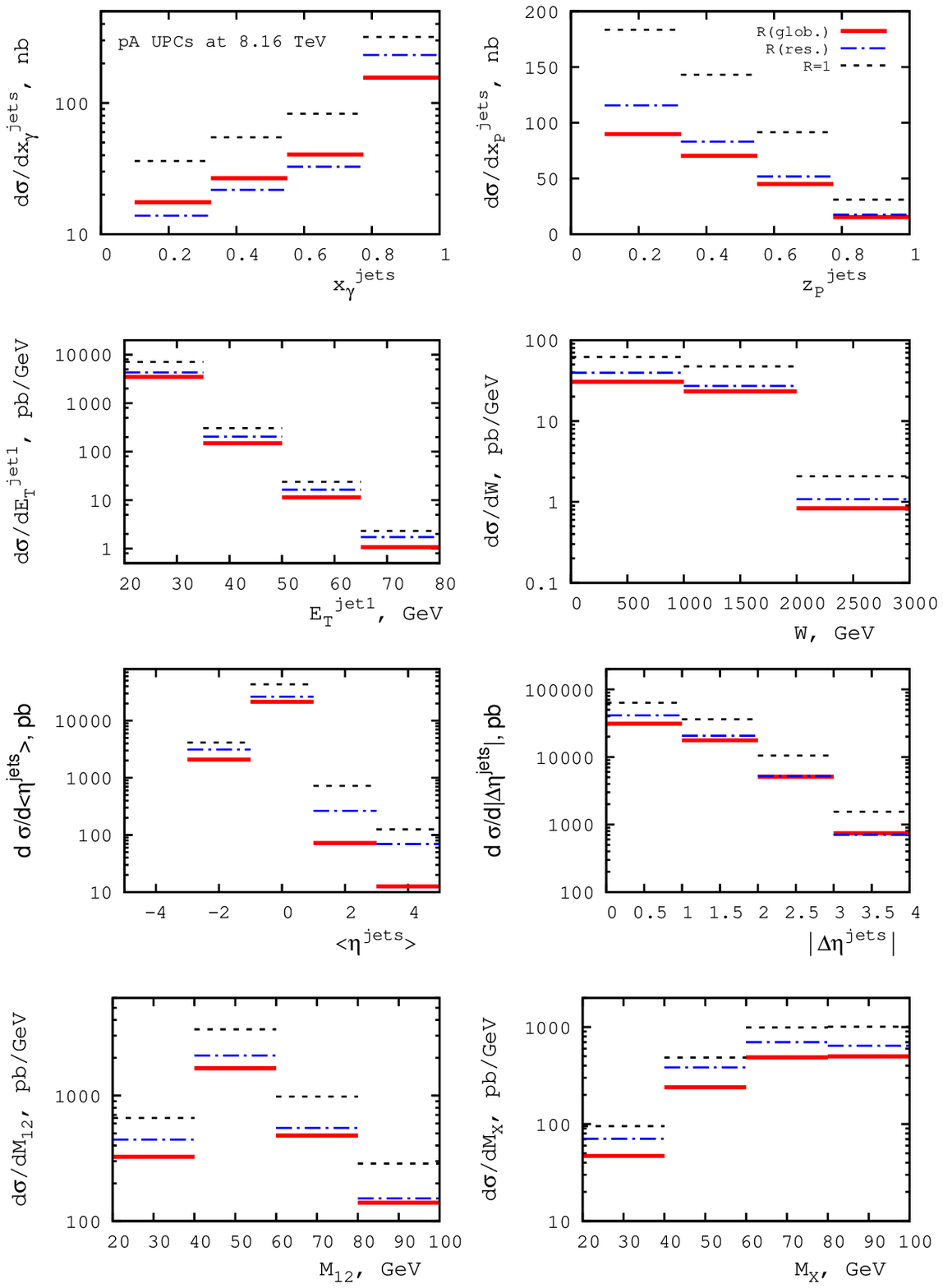,scale=1.1}
 \caption{The same as in Fig.~\ref{fig:pA_fb}, but at $\sqrt{s_{NN}}=8.16$ TeV.}
 \label{fig:pA88_fb}
\end{center}
\end{figure}

\begin{figure}[t]
\begin{center}
\epsfig{file=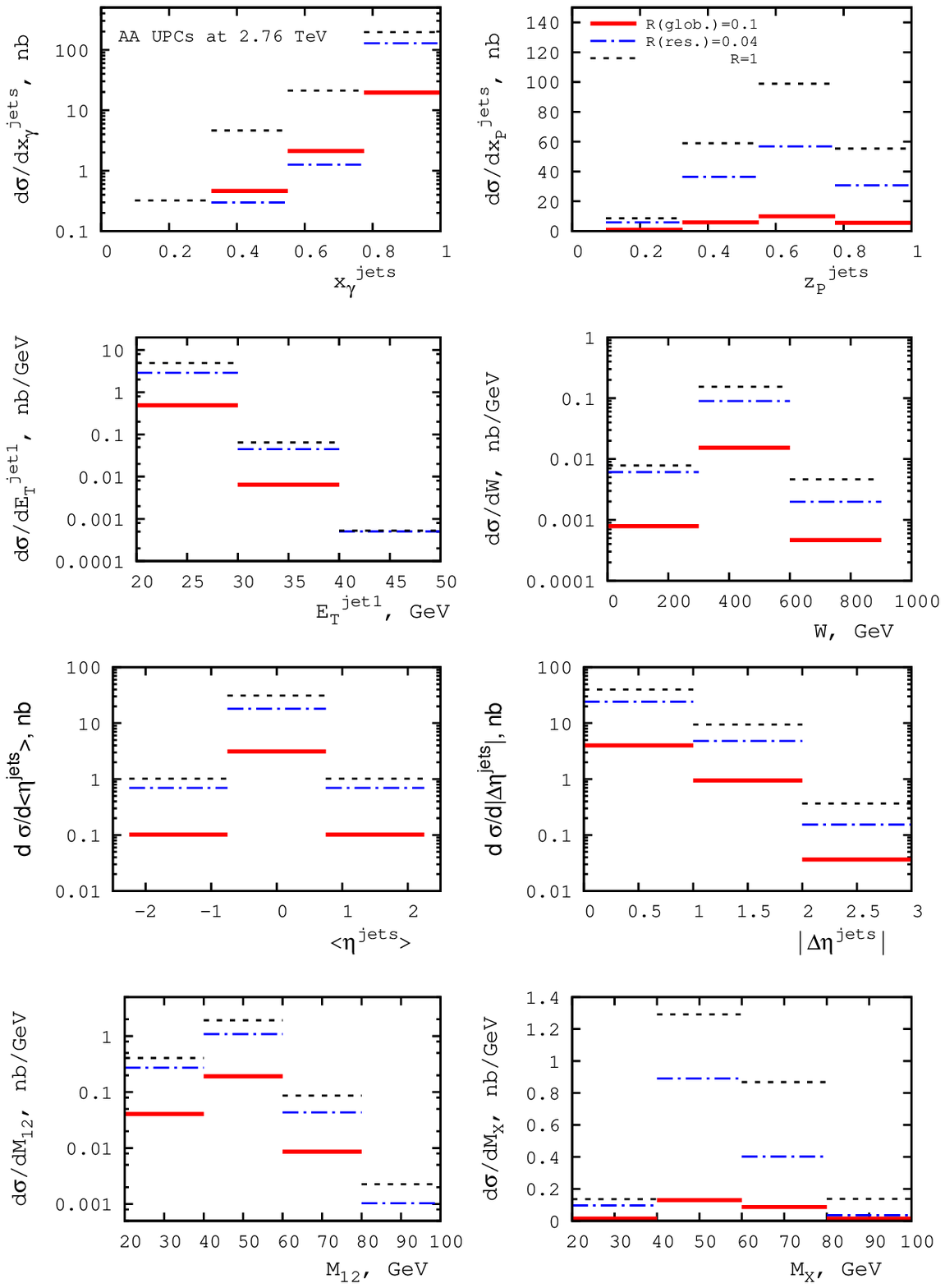,scale=1.1}
 \caption{The effect of diffractive factorization breaking on the differential cross section of diffractive photoproduction of dijets $d \sigma(AA\to A+2{\rm jets}+ X^{\prime}+A)$
 in $AA$ UPCs at $\sqrt{s_{NN}}=2.76$ TeV.}
 \label{fig:AA_fb}
\end{center}
\end{figure}

\begin{figure}[t]
\begin{center}
\epsfig{file=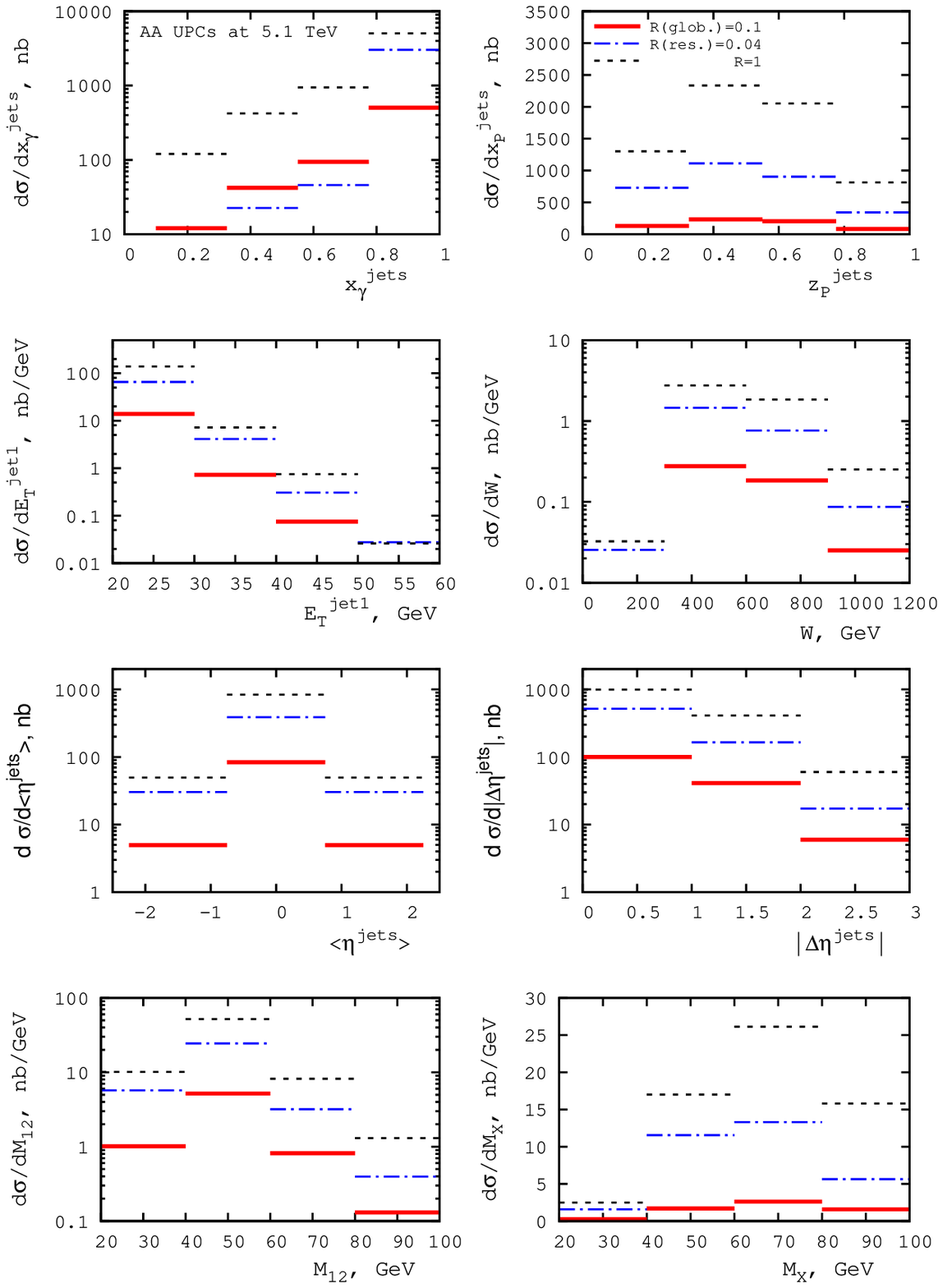,scale=1.1}
 \caption{The same as Fig.~\ref{fig:AA_fb}, but at $\sqrt{s_{NN}}=5.1$ TeV.}
 \label{fig:AA5_fb}
\end{center}
\end{figure}

The resulting cross sections of diffractive dijet photoproduction in $pp$, $pA$ and $AA$ UPC are presented in 
Fig.~\ref{fig:pp_fb}--\ref{fig:AA5_fb}. The red solid lines correspond to the global suppression factor of 
$R(\rm glob.)=0.5$ for the proton target and $R(\rm glob.)=0.1$ for the nucleus target (note that
in the case of $pA$ UPCs, we encounter a mixed situation);
the blue dot-dashed lines correspond to the suppression of the resolved photon contribution only:
$R({\rm res.})=0.4$ for the diffracting proton ($pp$ and the photon-from-nucleus contribution to $pA$) and 
$R({\rm res.})=0.04$ for the diffracting nucleus (the photon-from-proton contribution to $pA$ and $AA$).
For comparison, we also show our results that do not include the effect of diffractive QCD factorization
breaking by black dotted lines labeled ``$R=1$". Note that in all cases, we show only the predictions corresponding
to the central value of the renormalization and factorization scale $\mu=E_{T}^{\rm jet 1}$.

As one observes, the most sensitive variable to distinguish global from resolved-only suppression is 
$x_\gamma^{\rm jets}$
as expected, where resolved-only suppression is smaller in the highest and larger in the lower bins. As also
observed previously in diffractive dijet photoproduction at HERA~\cite{Klasen:2005dq}, the distributions in
$E_T^{\rm jet1}$ also show differences, i.e.\ resolved-only suppression results in harder spectra than global
suppression. These differences are more pronounced in $pA$ collisions compared to 
$pp$ collisions due to
the enhanced photon flux and asymmetric experimental setup. In $pA$ UPCs, also the $z_{\Pomeron}^{\rm jets}$ distributions
differ, i.e.\ resolved-only suppression is less effective at small values of that variable, which are correlated
with large $x_\gamma^{\rm jets}$. Naturally, similar differences are then observed in the average rapidity and rapidity
difference distributions, from which the observed momentum fraction variables are derived. In $pA$ UPCs, the
differences of the two suppression schemes are furthermore enhanced at higher center-of-mass energy, where the
low $z_{\Pomeron}^{\rm jets}$ region is particularly enhanced. 
In contrast, 
in $AA$ collisions the shape of the
$z_{\Pomeron}$ distribution is quite different from those in $pp$ and $pA$ collisions,  
but is similar in the two used suppression schemes, which makes it less sensitive to the factorization breaking pattern.

\newpage 

\newpage

\section{Conclusions}
\label{sec:discussion}

For the first time, using NLO pQCD, we make predictions for the cross sections of diffractive dijet photoproduction
in $pp$, $pA$ and $AA$ UPCs in the kinematics of Runs 1 and 2 at the LHC. 
Using general kinematic conditions and 
cuts on the final state, we found that the values of the cross section as a function of various variables are sufficiently large, 
i.e., this process can be observed. Compared to studies of this process in $ep$ scattering at HERA, we observe that 
UPCs provide an enhanced sensitivity to the low-$z_{\Pomeron}^{\rm jets}$ region probing the quark and gluon 
diffractive parton distributions in the proton and nuclei at small 
momentum fractions 
$z$ and an access to much larger values of $W$.

In our calculations, we used nuclear diffractive PDFs, which are strongly suppressed by nuclear shadowing;
neglecting this effect, our predictions for $AA$ UPCs and for the photon--nucleus contribution to $pA$ UPCs would 
be larger by the factor of seven.

\begin{figure}[h]
\begin{center}
\epsfig{file=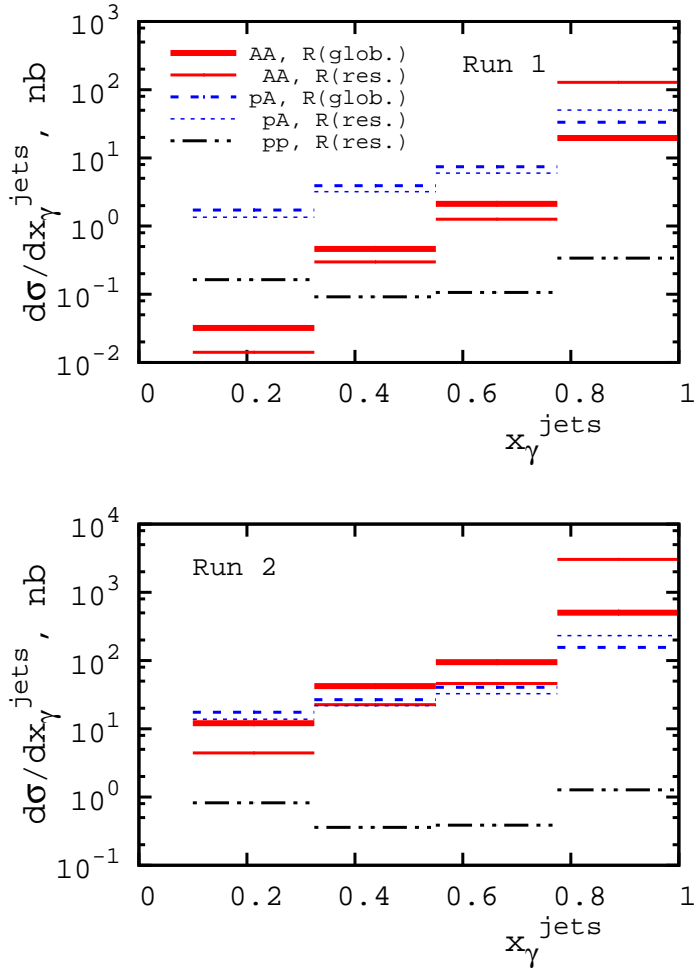,scale=1.5}
 \caption{The $x_{\gamma}^{\rm jets}$ dependence of the cross section of diffractive dijet photoproduction in $pp$, $pA$ and
 $AA$ UPCs at the LHC during Run 1 (upper panel) and Run 2 (lower panel) calculated using the global (thick lines) 
 and resolved-only (thin lines) schemes of factorization breaking.}
\label{fig:Summary_new}
\end{center}
\end{figure}

UPCs also give a new handle on the issue of diffractive QCD factorization breaking through its $A$ dependence: while
the two competing schemes of factorization breaking based on the global and resolved-only suppression factors give
rather similar predictions for $pp$ UPCs (like in the case of $ep$ scattering at HERA), the two scenarios give
rather different predictions for $AA$ UPCs and to some extent for $pA$ UPCs.  The best observable to look for this effect is 
the $x_{\gamma}^{\rm jets}$ dependence at large $x_{\gamma}^{\rm jets}$, which is dominated by the direct photon contribution
and where the ordering between the cross sections calculated using $R(\rm glob.)$ and $R(\rm res.)$ changes.
This is illustrated in Fig.~\ref{fig:Summary_new} summarizing our results for the $x_{\gamma}^{\rm jets}$ dependence in Run 1
(upper panel) and Run 2 (lower panel). Note that in this figure, the two schemes of factorization breaking for the $pp$ case are
indistinguishable.

The results presented in this work are based on the NLO collinear factorization formalism of pQCD.
In the framework of high-energy QCD, 
diffractive dijet production in photon--proton and photon--nucleus collisions was considered in the Color Glass Condensate
(CGC) formalism at leading order in Ref.\ \cite{Altinoluk:2015dpi}. 
It was found that the effects of gluon saturation can be searched for 
in the dijet azimuthal angle correlations and $t$ distributions.
In addition, a theoretical framework for diffractive production of jets in the
QCD shock-wave approach has started to be developed in a series of papers~\cite{Boussarie:2015qet,Boussarie:2015acw}. It will be interesting to confront these different approaches with LHC data in the future.

\appendix
\section{Suppression factors used for calculations in this paper}

For convenience of the reader, we give in this Appendix simple parametrizations of the various suppression factors
used in our calculations in this paper.

1. The rapidity gap survival probability factor of $S^2(x)$ for $pp$ UPCs, which is given 
by Eq.~(\ref{eq:S2_pp}) and shown in Fig.~\ref{fig:S2_pp},
can be fitted to better than 5\% accuracy by the following simple form:
\begin{equation}
S^2(x)=\frac{0.85}{1+a x+b x^2 } \,,
\label{eq:S2_pp_fit}
\end{equation}
where $a=14$ and $b=1.4$ at $\sqrt{s_{NN}}=7$ TeV;
$a=15$ and $b=4.8$ at $\sqrt{s_{NN}}=13$ TeV.

2. To quantify the suppression of the photon flux of the proton in $pA$ UPCs due to the strong interaction, 
it is convenient to introduce the factor of $f^{\rm sup}_{p}(x)$ (see Eqs.~(\ref{eq:flux_Ap})
and (\ref{eq:Gamma_pA})):
\begin{equation}
f^{\rm sup}_{p}(x) \equiv  \frac{\int d^2 b \, \Gamma_{pA}(b) f_{\gamma/p}(x,b)}{f_{\gamma/p}(x)} =\frac{0.71}{1+260 x}\,.
\label{eq:f_sup_p}
\end{equation}
The last equality gives a simple fit, which reproduces the calculation of $f^{\rm sup}_{p}$ to better 
than 5\% accuracy. Note that $f^{\rm sup}_{p}(x)$ gives the ratio of the red solid and the blue dot-dashed curves in
the right panel of Fig.~\ref{fig:flux_pA_Dec2015}.
The fit of Eq.~(\ref{eq:f_sup_p}) is valid both at $\sqrt{s_{NN}}=5.02$ TeV and $\sqrt{s_{NN}}=8.16$ TeV, 
since $f^{\rm sup}_{p}(x)$
does not change in this energy interval to better than a fraction of a percent accuracy.

\acknowledgments
 
The authors would like to thank M.~Strikman and M.~Zhalov for carefully reading the manuscript and useful comments.
VG would like to thank W.~Vogelsang for useful discussions of photon PDFs 
and the
Institut f\"ur Theoretische Physik, Westf\"alische Wilhelms-Universit\"at M\"unster for hospitality.
The work of VG is partially supported
by a grant of Deutscher Akademischer Austauschdienst (DAAD). 
The work of MK is partially supported by the BMBF Verbundprojekt 05H2015 through grant 05H15PMCCA.

\end{document}